\documentclass[%
 reprint,
 amsmath,amssymb,
 aps,
]{revtex4-2}

\usepackage{graphicx}% Include figure files
\usepackage{subcaption}
\usepackage{dcolumn}% Align table columns on decimal point
\usepackage{makecell}
\usepackage{bm}% bold math
\begin{document}

\preprint{APS/123-QED}

\title{Scattered light noise at LIGO Livingston Observatory during O4}% Force line breaks with \\
%\thanks{A footnote to the article title}%

\author{Debasmita Nandi$^1$,  Anamaria Effler$^2$, Siddharth Soni$^3$, Tabata Aira Ferreira$^1$, Robert Schofield$^4$, Huyen Pham$^2$, Timothy O'Hanlon$^2$, V. V. Frolov$^2$, Gabriela Gonz\'alez$^1$}
\email{dnandi1@lsu.edu}
\affiliation{$^1$Department of Physics and Astronomy, Louisiana State University, 202 Nicholson Hall, Baton Rouge, LA 70803, USA \\
$^2$LIGO Livingston Observatory, Livingston, LA 70754, USA\\
$^3$University of California, Riverside, CA 92521, USA \\
$^4$Department of Physics, University of Oregon, Eugene, OR 97403, USA}%

% \author{Second Author}%
% \collaboration{MUSO Collaboration}%\noaffiliation

% \author{Charlie Author}
%  \homepage{http://www.Second.institution.edu/~Charlie.Author}
% \affiliation{
%  Second institution and/or address\\
%  This line break forced% with \\
% }%
% \affiliation{
%  Third institution, the second for Charlie Author
% }%
% \author{Delta Author}
% \affiliation{%
%  Authors' institution and/or address\\
%  This line break forced with \textbackslash\textbackslash
% }%

% \collaboration{CLEO Collaboration}%\noaffiliation

\date{\today}% It is always \today, today,
             %  but any date may be explicitly specified

\begin{abstract}
Scattered light is one of the most common sources of noise in the LIGO gravitational wave detectors. Light scattering is a highly non-linear process through which motion at low frequencies gets up-converted and creates noise in a higher frequency band in the detector data. From the beginning of the fourth observation run, many glitches appeared in the data of LIGO Livingston detector in the frequency range $10$-$40$ Hz, and the morphology of these glitches suggested that they were produced by scattered light. From our analysis, we identified two different populations of scattered light glitches, one group having higher SNR than the other. The glitches of the high-SNR group were solely modulated by microseismic ground motion (ground motion in $0.1$-$1.0$ Hz) and in this paper, we present models of possible coupling mechanisms for these glitches. We also present results of a statistical correlation analysis based on our models, which indicates that the microseismic ground motion at the corner station along the X direction is the one most correlated with the noise which create these high SNR glitches. After installing baffles very close to the test mass mirrors, we have noticed a significant reduction in the rate and SNR of these glitches. The low-SNR glitches were primarily modulated by high frequency ($10$-$30$ Hz) vertical ground motion at the corner station, and this motion was coupling through a specific vacuum chamber at the corner station. After installing an additional seismic isolation platform in that vacuum chamber, these glitches have disappeared. 
% \begin{description}
% \item[Usage]
% Secondary publications and information retrieval purposes.
% \item[Structure]
% You may use the \texttt{description} environment to structure your abstract;
% use the optional argument of the \verb+\item+ command to give the category of each item. 
% \end{description}
\end{abstract}

%\keywords{Suggested keywords}%Use showkeys class option if keyword
                              %display desired
\maketitle

%\tableofcontents

\section{\label{sec:intro}Introduction}
Since the first detection of gravitational waves from merging black holes on September 14, 2015 \cite{LIGOScientific:2016aoc}, the LIGO-Virgo-KAGRA collaborations have detected 218 gravitational wave signals from astrophysical sources till the first part of the fourth observation run (O4a: 2023 May 24 15:00:00 UTC to 2024 January 16 16:00:00 UTC)  \cite{LIGOScientific:2025slb}. These detectors are 3-4 km-long interferometers that can detect very small amplitude perturbations of space-time, with an arm length difference smaller than $10^{-18}$m in a frequency band between 10 Hz and 500 Hz. 

The gravitational wave signals that have been detected so far are transient in nature and their detection is limited by the different noise sources which reduce the sensitivity of the detectors. Noise can be of instrumental or environmental origin and can be stationary or transient in nature. Stationary noise sources are mainly attributed to quantum shot noise, radiation pressure noise, thermal noise, seismic noise, and control noise \cite{Capote:2024rmo}. Transient noise, or ``glitches'', appear due to various reasons \cite{LIGO:2024kkz, Glanzer:2022avx} and  can directly impact the astrophysical searches by mimicking an astrophysical signal or by overlapping with a signal, and can also lead to errors in parameter estimation \cite{Udall:2025bts}. This is why it is very important to characterize these noises, identify their sources, and mitigate them. In this paper, we focus on one of the major population of glitches caused by scattered light in the data of LIGO Livingston Observatory (LLO) during the fourth observation run (O4).  

Due to small imperfections on their surfaces, all test mass mirrors scatter a fraction of light incident on them away from the direction of the main beam. Sometimes this scattered light rejoins the main beam after being reflected back from some moving surface in the vicinity. The relative motion of that surface with respect to the test mass mirror adds a time-dependent phase fluctuation in the main beam \cite{Accadia:2010zzb, Canuel:2013suy, Ottaway:2012oce}, as shown in equation \ref{eq:phase_fluctuation}. This time-dependent phase fluctuation creates a phase noise in the strain data of the detector following equation \ref{eq:scatter_eq}. 

\begin{align}
    &\phi(t) = \phi_0 + \delta \phi_{sc}(t) = \frac{4 \pi}{\lambda}|x_0 + \delta x_{sc}(t)| \label{eq:phase_fluctuation} \\
    & h_{ph}(f) = A \frac{\lambda}{8 \pi L} \mathcal{F}[\sin \delta \phi (t)] 
    \label{eq:scatter_eq} 
\end{align}
Where $A$ is the fraction of the electric field that is being scattered, $\lambda$ is the wavelength of the laser ($1064$ nm in LIGO), $L$ is the length of the interferometer arm ($4$ km), $x_0$ is the distance between the moving surface and the test mass mirror, $\delta x_{sc}(t)$ is the time-dependent relative displacement of the moving surface which adds  a time-dependent phase fluctuation $\delta \phi_{sc} (t)$ to the static phase $\phi_0$; $\mathcal{F}$ denotes the Fourier transform. 

If this time-dependent displacement $\delta x(t)$ is much smaller than the wavelength $\lambda$, then small-angle approximation can be used for the $sin (\delta \phi (t))$ term. But if $\delta x(t)$ is comparable to $\lambda$, then the series expansion of $sin (\delta \phi (t))$ gives some non-linear terms. This non-linearity ``upconverts" the low-frequency motion  and creates noise at higher frequencies in the primary gravitational wave channel. In the time-frequency spectrogram, this noise appears as arches \cite{LIGO:2020zwl}.
The time separation between the consecutive arches gives us information about the frequency with which the scattering surface is moving. The peak frequency of the arches is determined by the velocity of the moving surface as given by equation \ref{eq:freq_prediction}:
\begin{equation}
    f_{fringe}(t) = \left| \frac{2nv_{sc}(t)}{\lambda}\right|
    \label{eq:freq_prediction}
\end{equation}
 $v_{sc}(t)$ is the velocity of the surface with respect to the test mass mirror and $n$ is the number of times the scattered light gets reflected back and forth between the scatterer and the test mass mirror before joining the main beam. Multiple reflections of the same scattered light appear as higher harmonics of the same arch in the time-frequency spectrogram of the data. 
 
Most of the scattered light noise that appeared during the third observation run (O3) was caused by the transmitted light being scattered off of the shiny gold electrodes of the end test masses' (ETM) reaction chains \cite{LIGO:2020zwl}; and by a resonance of the arm cavity baffles (ACBs) at the corner station and at the Y-arm end station \cite{Soni:2023kqq}. Both of these noises were mitigated by implementing some changes in the instrument. 

\begin{figure*}
        \centering
        \includegraphics[width=0.65\textwidth]{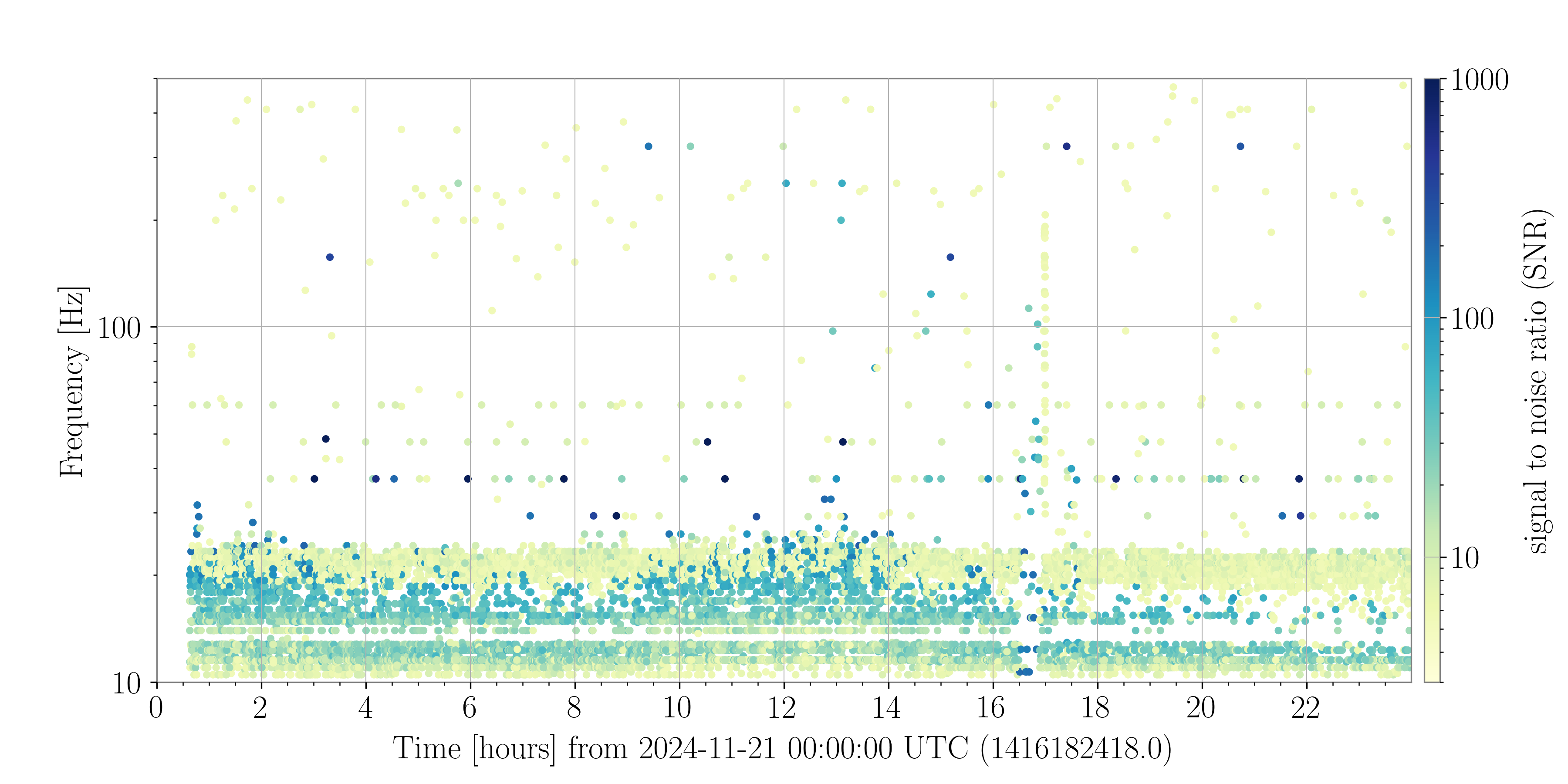}
        \caption{Omicron glitchgram of November $21$, $2024$. The X-axis denotes the UTC hours of the day and the Y-axis denotes frequency. Each dot in this image is a glitch identified by Omicron and the color represents the SNR of the corresponding glitch.}
        \label{fig:omicron}
    \end{figure*}
    
However, from the beginning of O4, many glitches appeared in LLO data during periods of elevated microseismic ground motion ($0.1$-$1.0$ Hz)~\cite{ferreira2025analysis}. These glitches had frequencies in the range $10$-$40$ Hz, and their time-frequency spectrograms revealed that they looked like arches, a characteristic morphology of glitches caused by scattered light. Here, we present a detailed discussion of these glitches. 
 
 This paper is organized as follows. In section \ref{section2}, we describe the properties of these glitches and identify them as two separate groups; in section \ref{model} and section \ref{investigations}, we present some models for possible noise coupling mechanisms and instrumental investigations for one of the groups of glitches; in section \ref{sec:20_Hz_glitches}, we describe the glitches from the second group; in section \ref{sec:instrumental_upgrades}, we describe the instrumental upgrades and their impact on the scattered light noise; in section \ref{sec:LHO}, we discuss the possibility of scattered light noise arising from similar coupling mechanisms at the LIGO Hanford Observatory (LHO); and in the final section (section \ref{conclusions}) we present conclusions and prospects for future work. 

 \section{Correlation of the noise with microseismic ground motion}\label{section2}
%\begin{itemize}
%    \item intro about omega scan and omicron
%    \item high microseism creates more scattering glitches
%    \item not having any fixed frequency
%\end{itemize}
To characterize the properties of the noise, we used tools that provide detailed insight about glitches. Two such tools are \textit{Omicron} and \textit{gwdetchar omega}. 

 \textit{Omicron} is an algorithm developed to find excess power in strain data using a method called $Q$ transform \cite{Robinet:2020lbf}. 
 % This is a modification of the usual Fourier transform which projects the time domain data onto a basis of sinusoidal Gaussian functions and extract the time-frequency information of the data. 
Excess power is measured by the signal-to-noise ratio (SNR). Omicron assigns some properties like time, duration, frequency, SNR, amplitude, and phase for all of the glitches it identified from the data. Figure \ref{fig:omicron} displays an Omicron glitchgram of LLO on November 21, 2024. Analyzing Figure~\ref{fig:omicron}, two distinct groups of glitches below 35~Hz can be identified: one with low SNR, marked by a light green color and persistent throughout the day; and another with higher SNR (dark green), exhibiting oscillations over a broader frequency range and increased rates at the beginning and middle of the day, then decreasing toward the end. To facilitate the analysis and explore potential correlations with microseismic ground motion, we divided the glitches into two categories---low and high SNR. In this section, we focus on the high-SNR group.

\begin{figure*}
    \centering
    \begin{minipage}[t]{0.7\textwidth}
        \vspace{0pt}
        \centering
        \includegraphics[width=\linewidth]{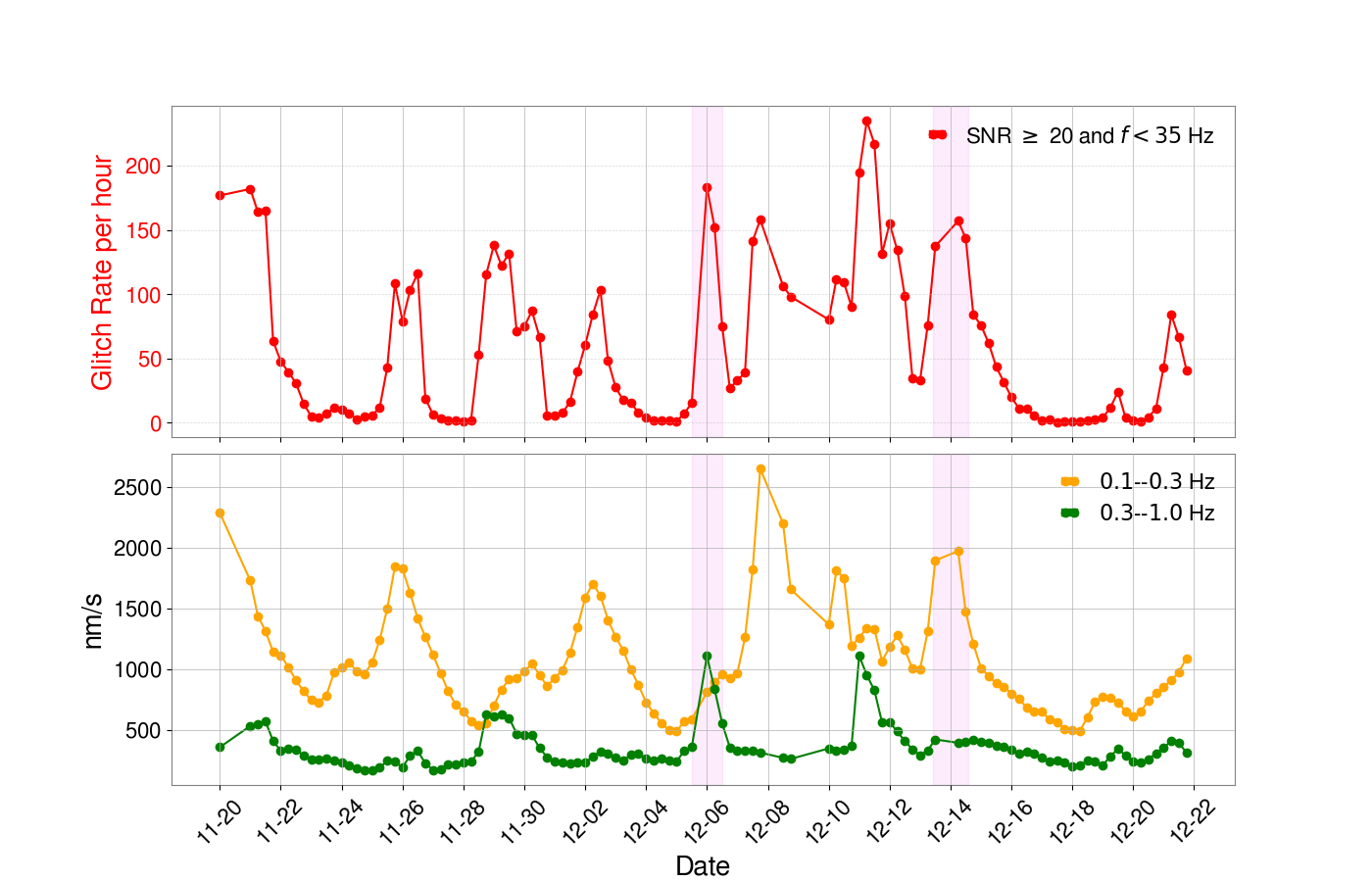}
    \end{minipage}%
    \hfill
    \begin{minipage}[t]{0.3\textwidth}
        \vspace{20pt}
        \centering
        \includegraphics[width=\linewidth]{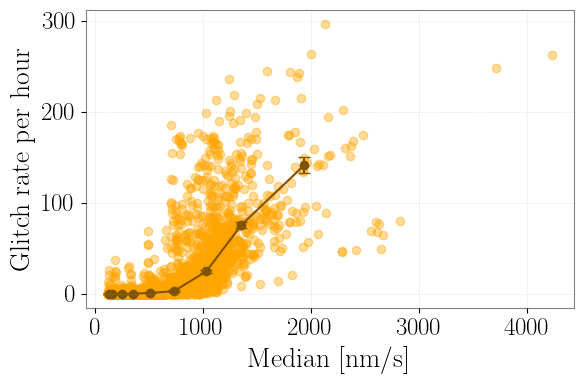} \\
        \vspace{2mm}
        \includegraphics[width=\linewidth]{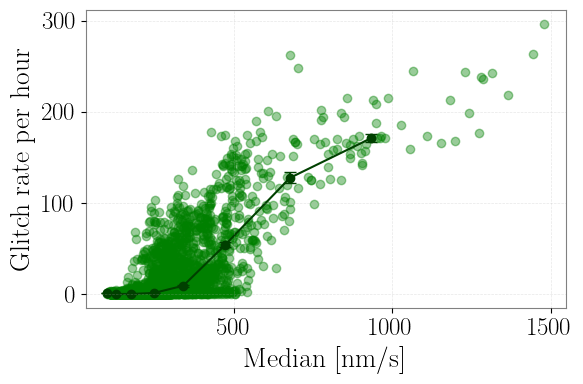}
    \end{minipage}
    \caption{\textit{Left:} Glitch rate per hour for high-SNR glitches as a function of the amplitude of the lower microseismic band (0.1–0.3 Hz) and the higher microseismic band (0.3–1 Hz), showing that this group appears to be independently correlated with both bands. \textit{Right:} Scatter plots of the hourly glitch rate as a function of ground motion amplitude in the lower band (orange, top) and higher band (green, bottom).}
    \label{fig:high_snr_glitches_vs_microseismic}
\end{figure*}

Figure~\ref{fig:high_snr_glitches_vs_microseismic} (left) shows, in red, the hourly rate of these high-SNR glitches over approximately one month of data (from November 19 to December 23, 2024). The bottom panel displays the amplitude of ground motion in the two microseismic bands over the same period: the lower band (0.1–0.3~Hz), shown in orange, and the higher band (0.3–1.0~Hz), shown in green. For this plot, we include all glitches with frequencies below 35~Hz and SNR~$\geq$~20. The glitch rate was calculated over 6-hour intervals and the ground motion was calculated as the median amplitude in each band-limited signal within the same 6-hour windows.

We observe that the glitch rate generally follows the trend of ground motion, with peaks modulated by both microseismic bands. For instance, on December 14 (highlighted by the second pink-shaded region), the amplitude in the higher band was low, while the lower band showed a pronounced peak - yet the glitch rate was high. Conversely, on December 6 (first pink region), the higher band exhibited significantly higher amplitude than the lower band, and the glitch rate was even slightly higher than in the previous case. We quantify the correlation between the glitch rate and ground motion using the Spearman correlation coefficients computed from the raw data. For the lower microseismic band ($0.1$-$0.3$ Hz), we found a strong and statistically significant correlation, with a Spearman coefficient of $0.73$ with p-value $< 10^{-16}$. A similarly strong correlation is observed for the higher microseismic band, with Spearman coefficients of $0.64$ and p-value $<10^{-16}$. These examples suggest a correlation between the high-SNR glitches and microseismic motion in both bands, with each band potentially contributing independently to glitch generation. 

To support this observation, the right side of Figure~\ref{fig:high_snr_glitches_vs_microseismic} shows scatter plots of the hourly glitch rate as a function of ground motion amplitude in the lower band (top) and higher band (bottom) covering the entire first part of the fourth observing run (O4a). A clear increase in glitch rate is evident with increasing ground motion in both bands. Here, we do not disentangle the times when only one of the bands is high or when both of them are high. The black curves represent the median glitch rate computed within amplitude bins, providing a smoothed view of the overall trend and reinforcing the correlation.

\begin{figure*}
    \centering
    \includegraphics[width=0.65\textwidth]{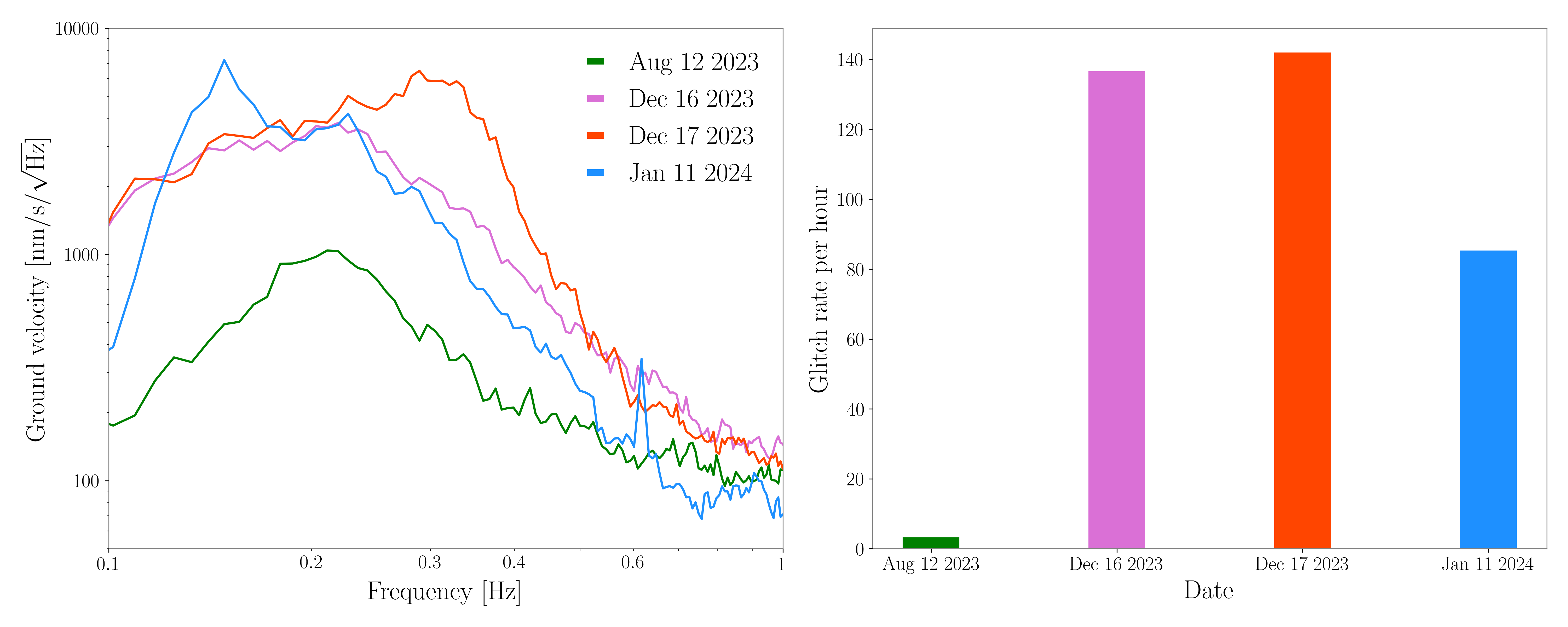}
    \caption{\textit{Left: }Amplitude spectral density of ground velocity, the green trace shows the ground motion during a quiet day, the other three traces show ground motion during the days having lot of glitches; \textit{Right: }The hourly rate of glitches, chosen from the frequency range $10$-$40$ Hz.}
    \label{fig:glitch_rate}
\end{figure*}

\begin{figure*}
    \centering
    \includegraphics[width=0.7\textwidth]{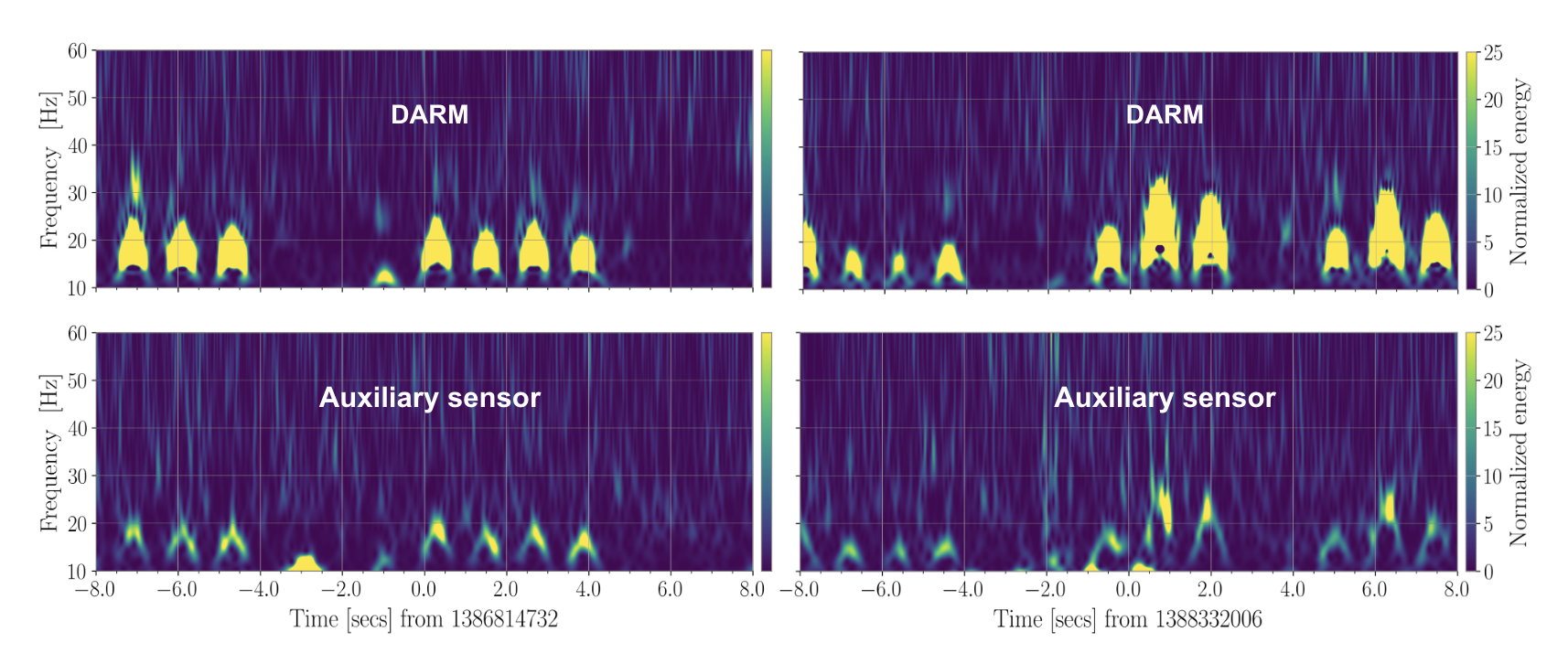}
    \caption{\textit{Top}: Scattered light glitches chosen from Nov $21$ $2024$ appearing in the gravitational wave channel (DARM: differential arm length); \textit{Bottom}: The same glitches appearing in a photodiode  (LSC-POP-9Q) which is witnessing the noise.}
    \label{fig:scatter_darm_pop9q}
\end{figure*}

During O3, a group of glitches created by scattered light was correlated with the dominant microseismic frequency in the control drive applied to the main chain by the reaction chain of the End Test Mass mirrors (ETMs) \cite{LIGO:2020zwl} and another group was correlated with a structural resonance of the arm cavity baffle (ACB) at $1.6$ Hz \cite{Soni:2023kqq}. However, in O4, the scattered light noise did not seem to be correlated with any particular frequency. We observed the scattered light noise in the data as long as the ground motion in the band $0.1$-$0.5$ Hz was above a certain threshold. This can be seen more clearly in Figure \ref{fig:glitch_rate}, which shows the hourly glitch rate and the amplitude spectral density of the ground motion for four different days. During three of these days, ground motion between $0.1$-$0.5$ Hz was high, which led to a high rate of glitches in the $10$-$40$ Hz frequency band. It is important to note here that these three days had dominant ground motion in different frequencies between $0.1$-$0.5$ Hz. From this we understand that no specific frequency is  responsible for the increased glitch rate. Instead, if there is excess motion in any frequency between $0.1$-$0.5$ Hz, then it results in an increased rate of glitches.

\textit{ Gwdetchar omega} is a tool which is used to display the time-frequency spectrogram of the data recorded in the gravitational-wave channel and in the auxiliary sensors (this includes seismometers, temperature sensors, photodiodes, accelerometers etc \cite{AdvLIGO:2021oxw}). This tool is very useful for finding the auxiliary sensors which witness the same noise as the gravitational wave channel and may help localize the source of the noise. Figure \ref{fig:scatter_darm_pop9q} shows some of the scattered light glitches from November $21$, $2024$. These glitches appear as arches in the gravitational wave channel (DARM: differential arm length) and also in another photodiode (LSC-POP-9Q), which is an auxiliary sensor witnessing the noise. 

\begin{figure}
    \centering
    \includegraphics[width=0.5\textwidth]{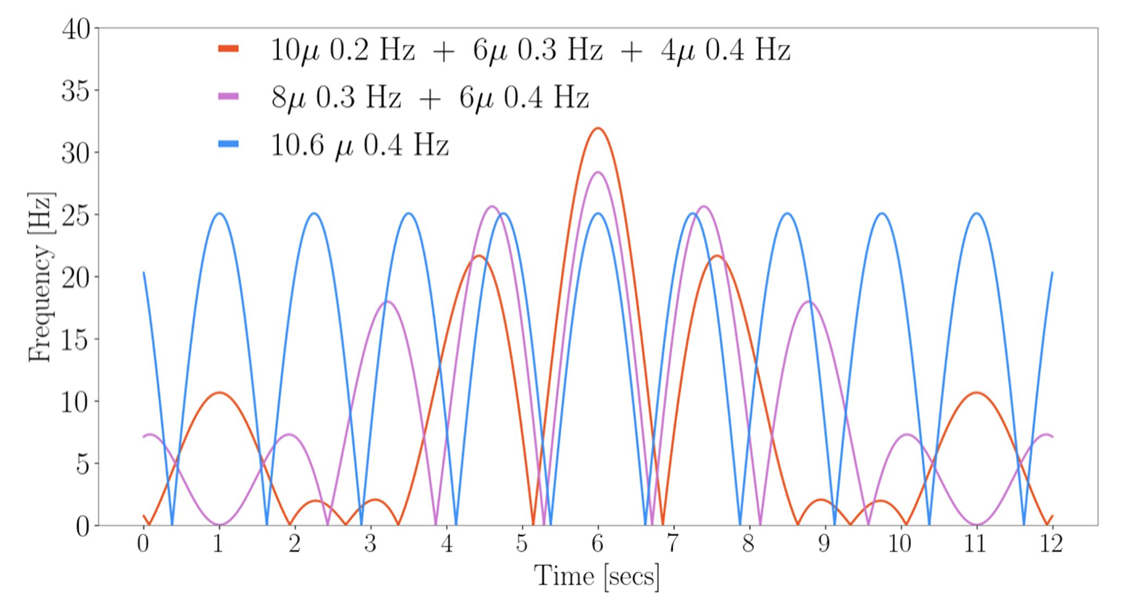}
    \caption{Predicted fringe frequencies by adding motion at different frequencies. The labels indicate the amplitudes and frequencies of motion which are added together.}
    \label{fig:add_motion_at_diff_freq}
\end{figure}

We have noticed that the peak frequencies of the scatter arches fall in the frequency range $10$-$40$ Hz (see the top panel in figure \ref{fig:scatter_darm_pop9q}), which means that the velocity of the scattering surface can be anywhere between $5 \;\mu/s$ and $20 \;\mu/s$, following equation \ref{eq:freq_prediction}. We did not find any optics that had this large velocity at frequencies between $0.1$-$0.5$ Hz. But if we add motion at different frequencies all together, the resultant velocity can be higher than the velocity at any particular frequency. Figure \ref{fig:add_motion_at_diff_freq} illustrates an example of this, where we show that a higher peak frequency of the scatter arches can be obtained by adding motion at different frequencies even if the velocity at any particular frequency is not that high. It is possible that the seismic motions at different frequencies between $0.1$-$0.5$ Hz are also getting combined together like this to produce the noise.

\begin{figure*}
\centering
\begin{subfigure}{0.45\textwidth}
  \centering
  \includegraphics[width=1.0\textwidth]{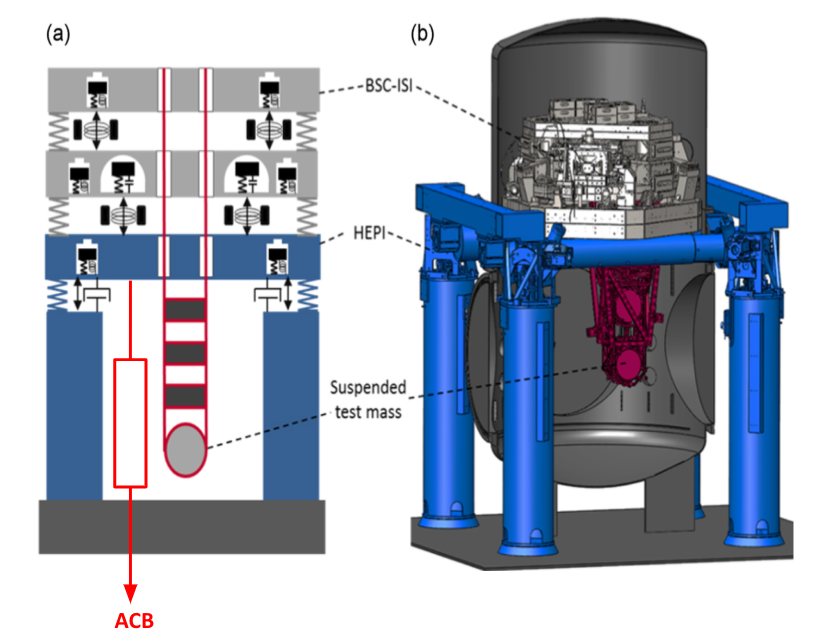}
  \caption{}
  \label{fig:sus_sei}
\end{subfigure}%
\begin{subfigure}{0.45\textwidth}
  \centering
  \includegraphics[width=1.0\textwidth]{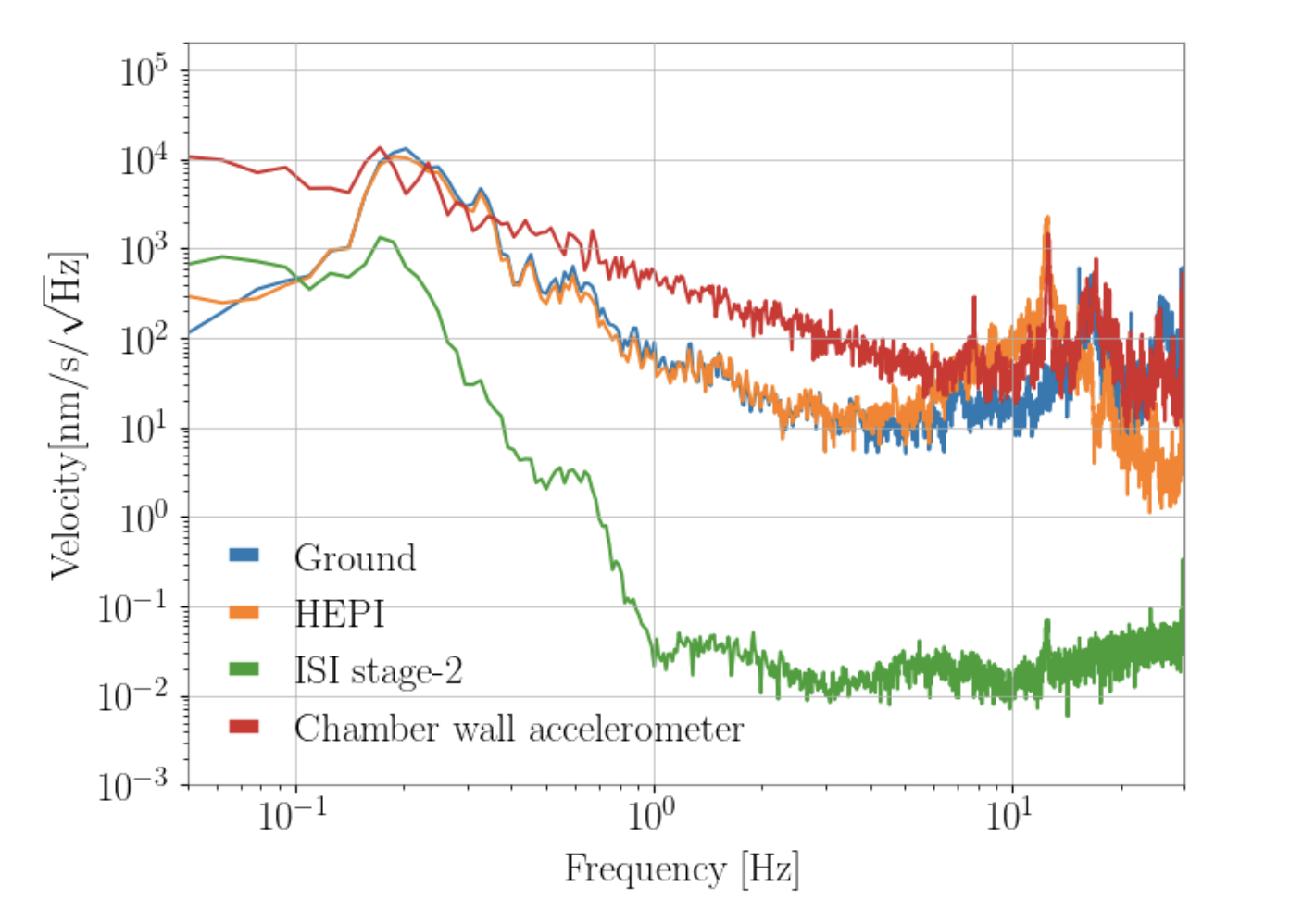}
  \caption{}
  \label{fig:sensor_spectra}
\end{subfigure}
\caption{(a) Schematic and CAD model of the seismic isolation systems supporting the core optics \cite{Matichard:2015eva}; the arm cavity baffle (ACB) is annotated; (b) Amplitude spectral density of motion recorded by different sensors.}
\label{fig:SEI_drawing_and_sensor_spectra}
\end{figure*}

 This makes us suspect that the potential source can be a surface that is not well isolated from the ground motion and lacks a sharp resonance at any particular frequency. As these frequencies are below the pendulum resonance frequency $0.45$ Hz, the seismic isolation system does not provide enough isolation. So, the relative motion between the scattering surface and the test mass is mostly modulated by the ground motion.

 \section{Modeling the noise} 
\label{model}

\begin{figure}
    \centering
    \includegraphics[width=0.5\textwidth]{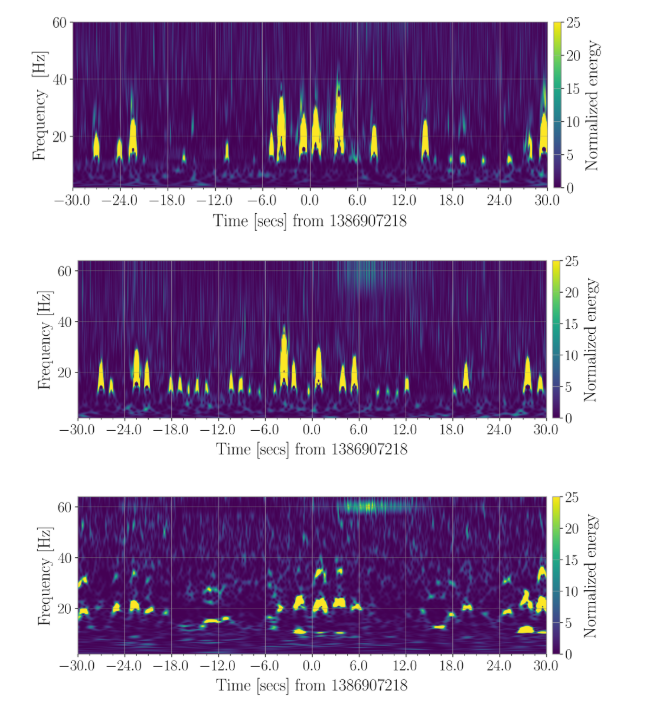}
    \caption{\textit{Top}: scattered light noise in DARM; \textit{Middle}: scattered light noise simulated by amplifying the motion; \textit{Bottom}: scattered light noise simulated without amplifying the motion.}
    \label{fig:simulated_noise}
\end{figure}

\begin{figure*}
    \centering
    \includegraphics[width=0.9\linewidth]{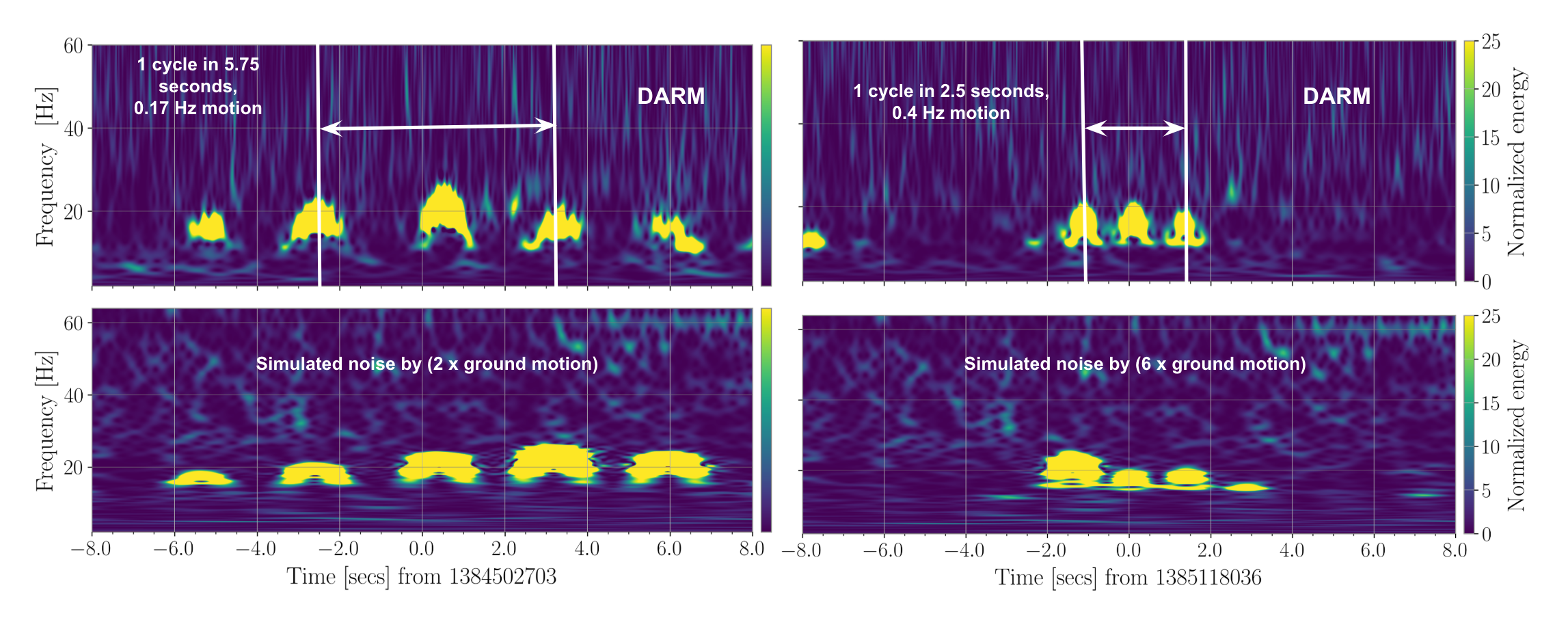}
    \caption{\textit{Top left:} Scattered light glitch in DARM, produced by 0.17 Hz motion of the scattering surface; \textit{Bottom left:} Scattered light glitch simulated by amplifying the ground motion by a factor of $2$; \textit{Top right:} Scattered light glitch in DARM produced by $0.4$ Hz motion of the scattering surface; \textit{Bottom right:} Scattered light glitch simulated by amplifying the ground motion by a factor of $6$.}
    \label{fig:amplify_ground_motion}
\end{figure*}

In order to model the noise, we needed to model the motion of the scattering surface ($\delta x_{sc}(t)$) and use that as input to equation \ref{eq:scatter_eq}. For the scattered light noise of O4, we developed two different models which explain two different possible coupling mechanisms. For the first model (sec. \ref{sec:3.1}), we use the unfiltered motion as recorded by the sensors of the primary seismic isolation platform (HEPI). For the second model (sec. \ref{sec:3.2}), we filter the motion of the HEPI platform by a transfer function which amplifies the motion at microseismic frequencies but suppresses the motion above $1$ Hz. 

Figure \ref{fig:sus_sei} shows a schematic of the suspension and seismic isolation system, which supports the test mass mirrors. The HEPI system provides the first stage of isolation from the ground motion. The next stage of isolation is provided by the Internal Seismic Isolation (ISI) platforms, which are inside the vacuum chambers. There are three stages in the ISI: stage $0$ is attached to the HEPI, stage 1 is suspended from  stage $0$, and stage $2$ is suspended from stage $1$. The final stage of the ISI (stage $2$) is the one which holds the optics table from which the test mass mirrors are suspended with quadruple pendulums and the beam splitter (BS) is suspended using a triple suspension. A baffle (Arm Cavity Baffle (ACB)), suspended from the HEPI in front of the test mass mirrors towards the main beam side, helps to capture any scattered beam from the test mass mirrors and redirect it away from the main beam. Similarly, the beam splitter has elliptical baffles that hang from the HEPI along both sides of the beam splitter.

Figure \ref{fig:sensor_spectra} shows the amplitude spectral density of motion recorded by different sensors. The blue trace shows the ground motion as measured by the seismometer located at the corner station. The orange trace shows the motion recorded by the L4C sensor at the HEPI platform of Input Test Mass (ITM) at the X-arm and the green trace shows the spectra of the GS13 sensor which measures the motion of the 2nd stage of ISI platform. The red trace shows the motion of the vacuum chamber which hosts the ITMX mirror and this is obtained from the accelerometer which is attached to the chamber wall. It is important to note here that the accelerometer is not very sensitive below $8$ Hz and hence the spectra has a $1/f$ signature which shows just the sensor noise at low frequencies. The peak at $0.2$ Hz in the ground and HEPI motion is due to elevated microseismic motion. The same peak is visible in the spectra of ISI stage-2 but has a much lower amplitude since this platform is suppressing the ground motion. The peaks in the spectra of the HEPI and the accelerometer around $10$ Hz are due to structural resonances.

\subsection{Modeling the noise using unfiltered motion}
\label{sec:3.1}
%\textcolor{violet}{AE:I think we need to give this a different name, because otherwise it just seems like the other is an over complication... the main difference is this model includes the higher frequency motion, which gets cut off by the pendulum. And then propagate that new name in the text further on.}
%In the previous model, the pendulum TF amplifies the low frequency motion (below $0.5$ Hz) but suppresses the high frequency motion (above $1$ Hz). So, we modeled the noise in another way in which the high frequency motion is included. 
For this model, we assumed the motion of the HEPI platform in the $0.05$-$30$ Hz frequency range to be the motion of the scattering surface and calculated the phase noise produced by that motion using equation \ref{eq:scatter_eq}. This model can explain the noise if scattering is occurring between the test mass mirrors and some structure which is moving as much as the seismic isolation platform that supports the vacuum chamber. The bottom panel of figure \ref{fig:simulated_noise} shows the Q-scan of a time series, where the phase noise is calculated using this model.

\subsection{Modeling the noise using filtered motion}
\label{sec:3.2}
 %There are some structures that are suspended from the seismic isolation platforms inside the vacuum chambers. In this model, we calculate the phase noise by modeling the motion of these structures.
 
\begin{figure}
    \centering
    \includegraphics[width=0.5\textwidth]{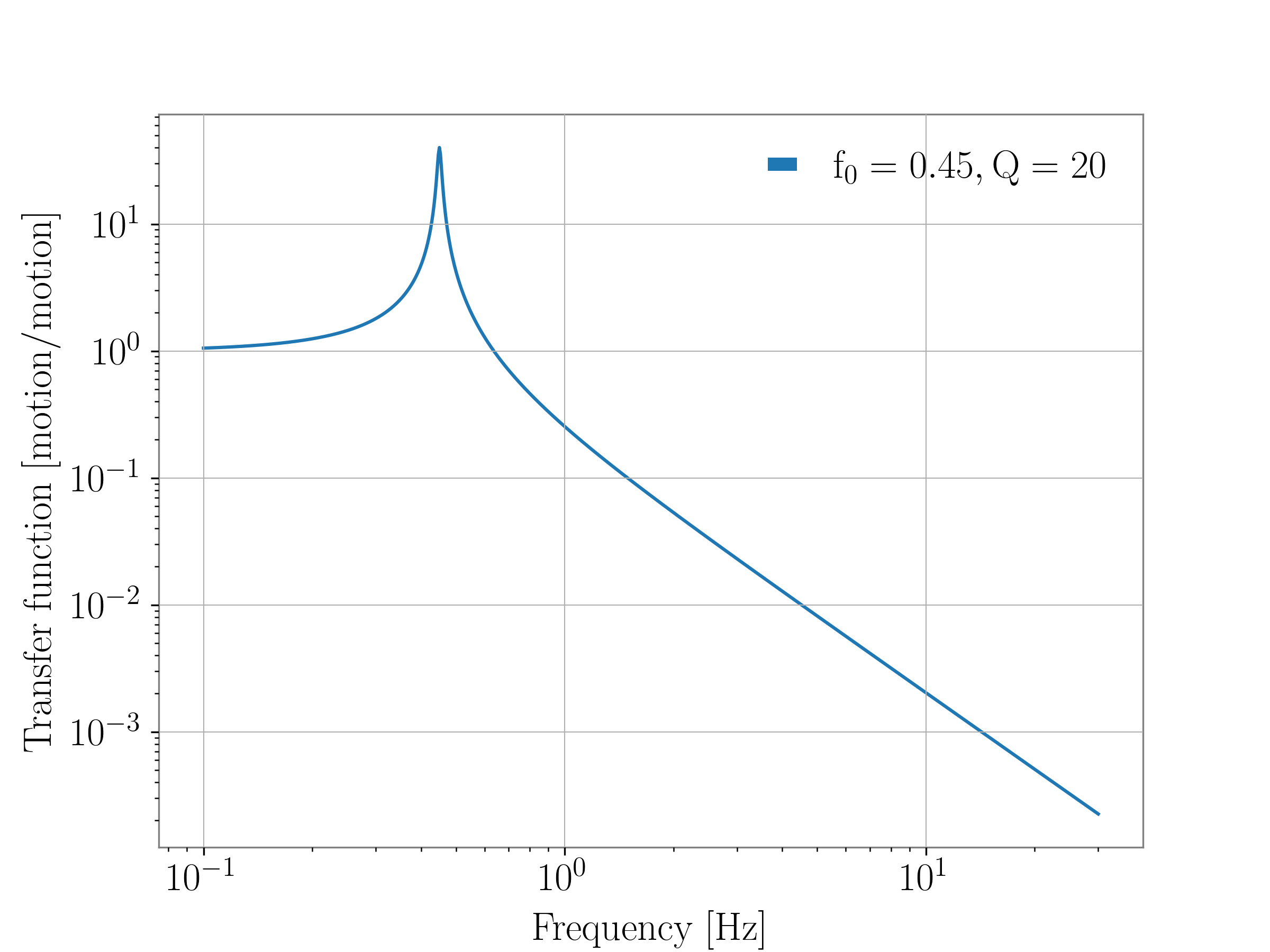}
    \caption{The transfer function of the scattering surface modeled as a damped driven harmonic oscillator having resonance frequency at $0.45$ Hz and a quality factor of $20$.}
    \label{fig:model_TF}
\end{figure}

 The motivation behind this model is based on the observation that the noise seen in the primary gravitational wave channel (DARM) could be reproduced by amplifying the microseismic ground motion ($0.1$-$0.5$ Hz) by a factor of $2$-$6$. This much amplification is not very difficult to obtain from some of the structures suspended from the seismic isolation platforms as a pendulum and which may have resonances at these low frequencies (below $0.5$ Hz). The amplification factor also varies based on the frequency of motion of the scattering surface. If the  noise is produced by a motion at lower frequencies (around $0.2$ Hz), then we need to amplify the ground motion by a factor of $1.5$-$2.0$ to match the level of noise seen in DARM. But if the scattering is due to a motion at slightly higher frequencies (around $0.4$ Hz), then we need to amplify the ground motion by a larger number, which varies between $4$-$6$. Figure \ref{fig:amplify_ground_motion} illustrates an example of this where the spectrograms in the top panel are of actual glitches in DARM and the spectrograms in the bottom panel are those calculated by amplifying the ground motion by some factor. This indicates that the scattering surface may have a transfer function which increases with frequency in the range $0.1$-$0.5$ Hz, which means it amplifies the ground motion more at $0.4$ Hz than it does at $0.2$ Hz. Based on this information, we modeled the transfer function ($TF$) of the scattering surface as a damped driven harmonic oscillator given by equation \ref{eq:model_TF}:
\begin{align}
    %TF(\omega) & = \frac{\omega_0^2}{\sqrt{m^2(\omega_0^2-\omega^2)^2+(\gamma \omega_0^2)^2}}
    TF(\omega) = \frac{\omega_0^2}{m^2(\omega_0^2-\omega^2)+i\omega_0\omega/2Q}
    \label{eq:model_TF}
\end{align}
where $\omega_0$ is resonance frequency of the pendulum ($0.45$ Hz in our case), $m$ is the mass of the pendulum, and $\gamma$ is the damping coefficient given as
\begin{align}
    \gamma = \frac{1}{2Q}
\end{align}
where $Q$ is the quality factor, which determines the nature of the resonance. The higher the $Q$ value, the sharper the resonance at the resonance frequency. Since the scattering surface does not have any specific resonance frequency and some of the known sharp resonances were mitigated in O3 \cite{Soni:2023kqq}, we used a low $Q$ value ($Q=20$) in our model. Figure \ref{fig:model_TF} shows the transfer function modeled using equation \ref{eq:model_TF}. 

\begin{figure*}
\centering
\begin{subfigure}{.45\textwidth}
  \centering
  \includegraphics[width=1.0\textwidth]{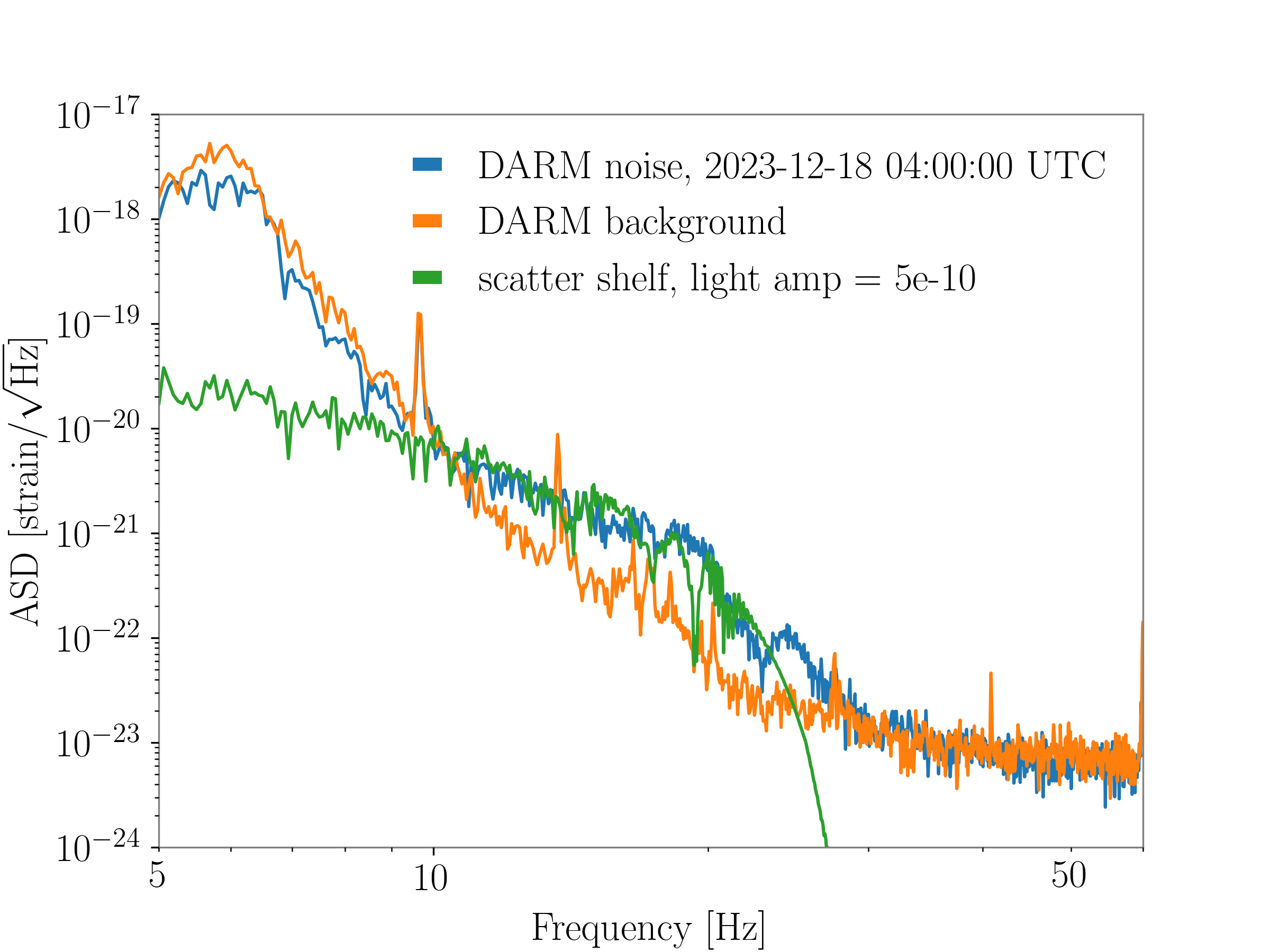}
  \caption{}
  \label{fig:shelf_with_amp_motion}
\end{subfigure}%
\begin{subfigure}{.45\textwidth}
  \centering
  \includegraphics[width=1.0\textwidth]{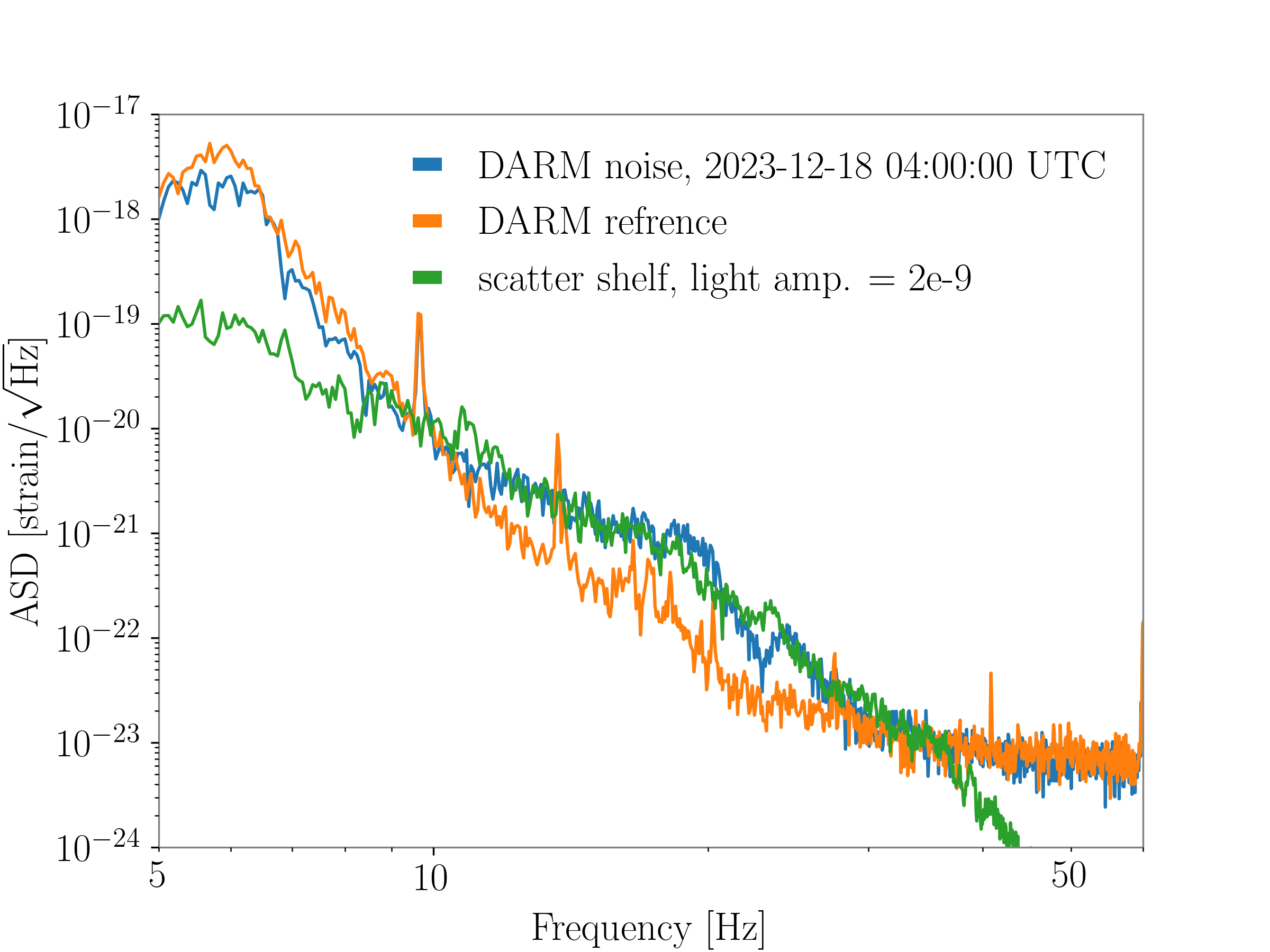}
  \caption{}
  \label{fig:shelf_wout_amp_motion}
\end{subfigure}
\caption{Simulated scatter shelf (green traces) plotted along with the actual scatter shelf (blue traces) observed in DARM at that time; (a) Scatter shelf calculated using filtered HEPI motion (b) Scatter shelf calculated using unfiltered HEPI motion.}
\label{fig:light_amplitude}
\end{figure*}

 We filtered the motion of the HEPI platforms in $0.05$-$30$ Hz by this transfer function (equation \ref{eq:model_TF}) and assumed that to be the motion of the scattering surface, $\delta x_{sc}(t)$ in equation \ref{eq:phase_fluctuation}. This replicates motion of any structure which is hanging from HEPI platforms as a pendulum. Next, we calculated the phase noise using equation \ref{eq:scatter_eq}. The middle panel of figure \ref{fig:simulated_noise} shows an example of phase noise simulated using this model.

\subsection{Light amplitude of the scattered beam}
\label{sec:3.3}

%\textcolor{red}{GG: I htink this paragraph fits better in the next section, modeling the noise.}
One of the important parameters for characterizing the noise is the amplitude of light that is being scattered. We present this light amplitude as a fraction of the main beam. The amount of power in the scattered beam can be obtained by squaring this fraction and multiplying it by the circulating power inside the arm cavities ($300$ kW for present advanced LIGO detectors). The amplitude spectral density of the noise looks like a shelf, the height of which is determined by the amount of light in the scattered beam, and the corner frequency of the shelf is determined by the motion of the scattering surface. The green traces in figure \ref{fig:light_amplitude} show such a scatter shelf, calculated using the expression of $h_{ph}(f)$ from equation \ref{eq:scatter_eq}. The green trace in figure \ref{fig:shelf_with_amp_motion} shows a scatter shelf which is calculated assuming the amplitude of the scattered beam to be a fraction of $2\times 10^{-10}$ of the main laser beam and the motion of the scattering surface to be the motion of the HEPI platform filtered by the pendulum transfer function as described in section \ref{sec:3.2}. The green trace in figure \ref{fig:shelf_wout_amp_motion} shows a scatter shelf calculated by assuming the amplitude of the scattered beam to be a fraction of $2\times 10^{-9}$ of the main laser beam and the motion of the scattering surface to be the unfiltered motion of the HEPI platform as described in section \ref{sec:3.1}. We have verified with several examples that, if we use the filtered motion of the HEPI platform, then the required light amplitude varies between $2$-$5\times10^{-10}$, whereas if we do not filter the motion, then the required light amplitude is almost $10$ times higher and usually varies between $1$-$2\times10^{-9}$. Again, it is important to emphasize here that these two models assume two different scatter paths and hence, the required light amplitude is different in the two cases. 

\subsection{Localizing the source by correlation analysis}
To localize the source of the noise, we used the motion recorded by the L4C sensors on the HEPI platform and bandpassed it in the frequency range $0.05$–$30$ Hz at the five different BSC chambers (vacuum chambers which host the beam splitter and the four test mass mirrors) and then calculated the Pearson correlation coefficients between this motion and the band-limited RMS (BLRMS) of DARM in the $15$–$40$ Hz frequency band. We performed the analysis separately for the days with a high amount of ground motion only in the $0.1$–$0.3$ Hz band (lms) and for the days with a high amount of motion in both the $0.1$–$0.3$ Hz and $0.3$–$1.0$ Hz bands (hms). The first group contains $610$ glitches from 9 different days of O4b, and the second group contains $1001$ glitches from 11 different days of O4b. The average correlation coefficients for both groups are presented in Table \ref{tab:corr_table}. We performed this correlation analysis using both models, unfiltered motion (section \ref{sec:3.1}) and filtered motion (section \ref{sec:3.2}). In both cases, the motion at the corner station BSCs along the X direction showed the highest average correlation with the DARM BLRMS.

\begin{table*}
    \centering
    \footnotesize
    \setlength{\tabcolsep}{1.3pt}
    \begin{tabular}{|c|c|c|c|c|}
    \hline
    {} & \multicolumn{2}{|c|}{only lms is high} & \multicolumn{2}{|c|}{both lms and hms are high} \\
    \hline
         Chamber & \makecell{Mean correlation \\ (using unfiltered motion)} & \makecell{Mean correlation \\ (using filtered motion)} & \makecell{Mean correlation\\ (using unfiltered motion)} & \makecell{Mean correlation\\ (using filtered motion)} \\
    \hline
        \textbf{BS$_X$} & \textbf{0.395} & \textbf{0.4} & \textbf{0.201} & \textbf{0.313} \\
    \hline 
        BS$_Y$ & 0.135 & 0.139 & 0.039 & 0.083 \\
    \hline
        \textbf{ITMX$_X$} & \textbf{0.363} & \textbf{0.371} & \textbf{0.194} & \textbf{0.304} \\ 
    \hline
        ITMX$_Y$ & 0.067 & 0.057 & 0.009 & 0.035 \\ 
    \hline
        \textbf{ITMY$_X$} & \textbf{0.358} & \textbf{0.337} & \textbf{0.128} & \textbf{0.246} \\
    \hline
        ITMY$_Y$ & 0.067 & 0.057 & 0.024 & 0.066\\ 
    \hline
        ETMX$_X$ & 0.089 & 0.088 & 0.031 & 0.024\\ 
    \hline
        ETMX$_Y$ & 0.017 & 0.013 & -0.004 & -0.002\\ 
    \hline
        ETMY$_X$ & 0.032 & 0.021 & 0.002 & 0.002\\
    \hline
        ETMY$_Y$ & 0.05 & 0.05 & 0.02 & 0.01\\ 
    \hline
    \end{tabular}
    \caption{Mean correlation values between the DARM blrms and the HEPI motion at different chambers. The subscripts in the names of the chamber indicate the direction of the motion. The motion along X direction in the corner station shows the highest average correlation.}
    \label{tab:corr_table}
\end{table*}

\begin{figure*}
\centering
\begin{subfigure}{.45\textwidth}
  \centering
  \includegraphics[width=1.0\textwidth]{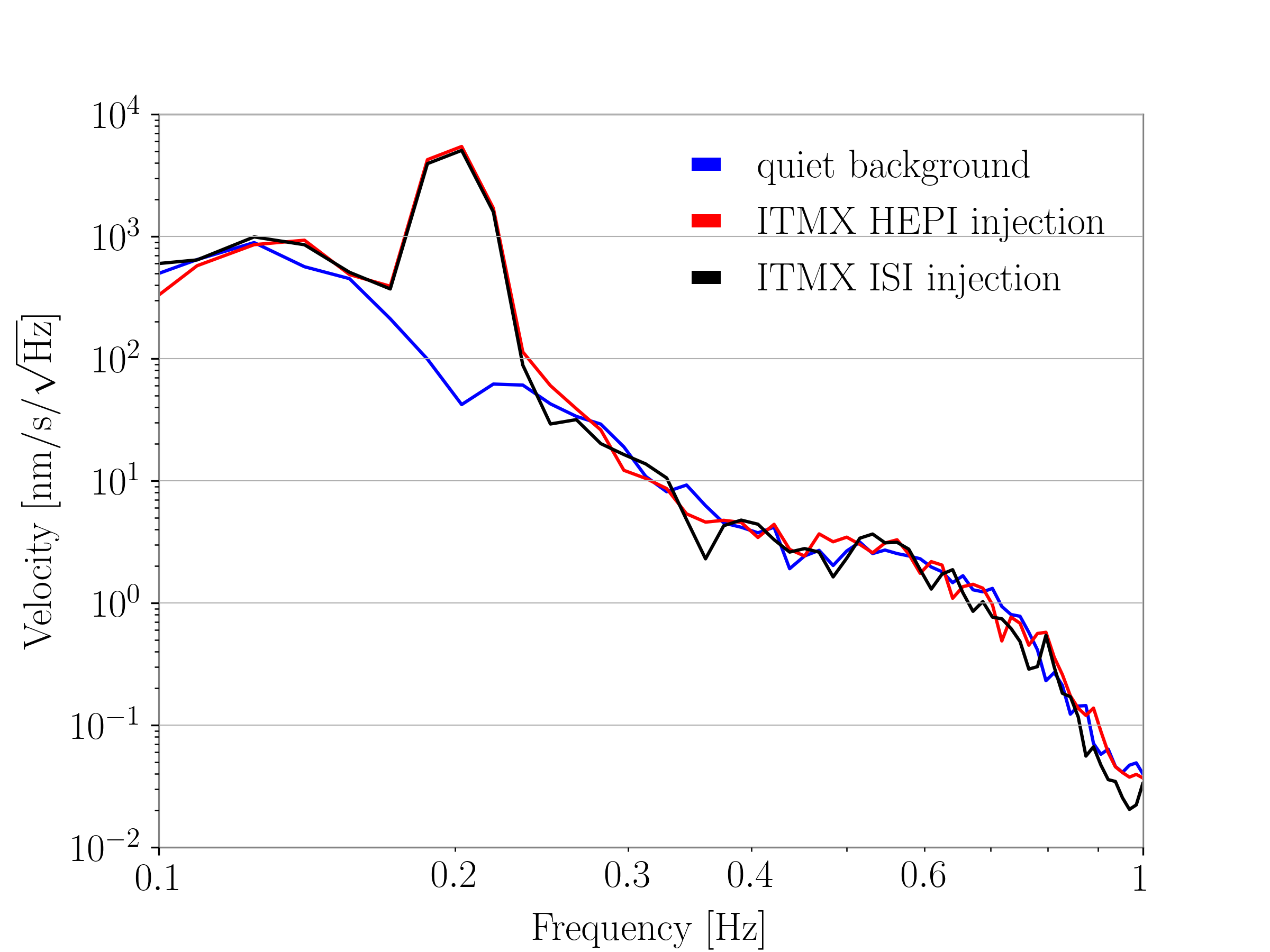}
  \caption{}
  \label{fig:itmx_injection_motion}
\end{subfigure}%
\begin{subfigure}{.45\textwidth}
  \centering
  \includegraphics[width=1.0\textwidth]{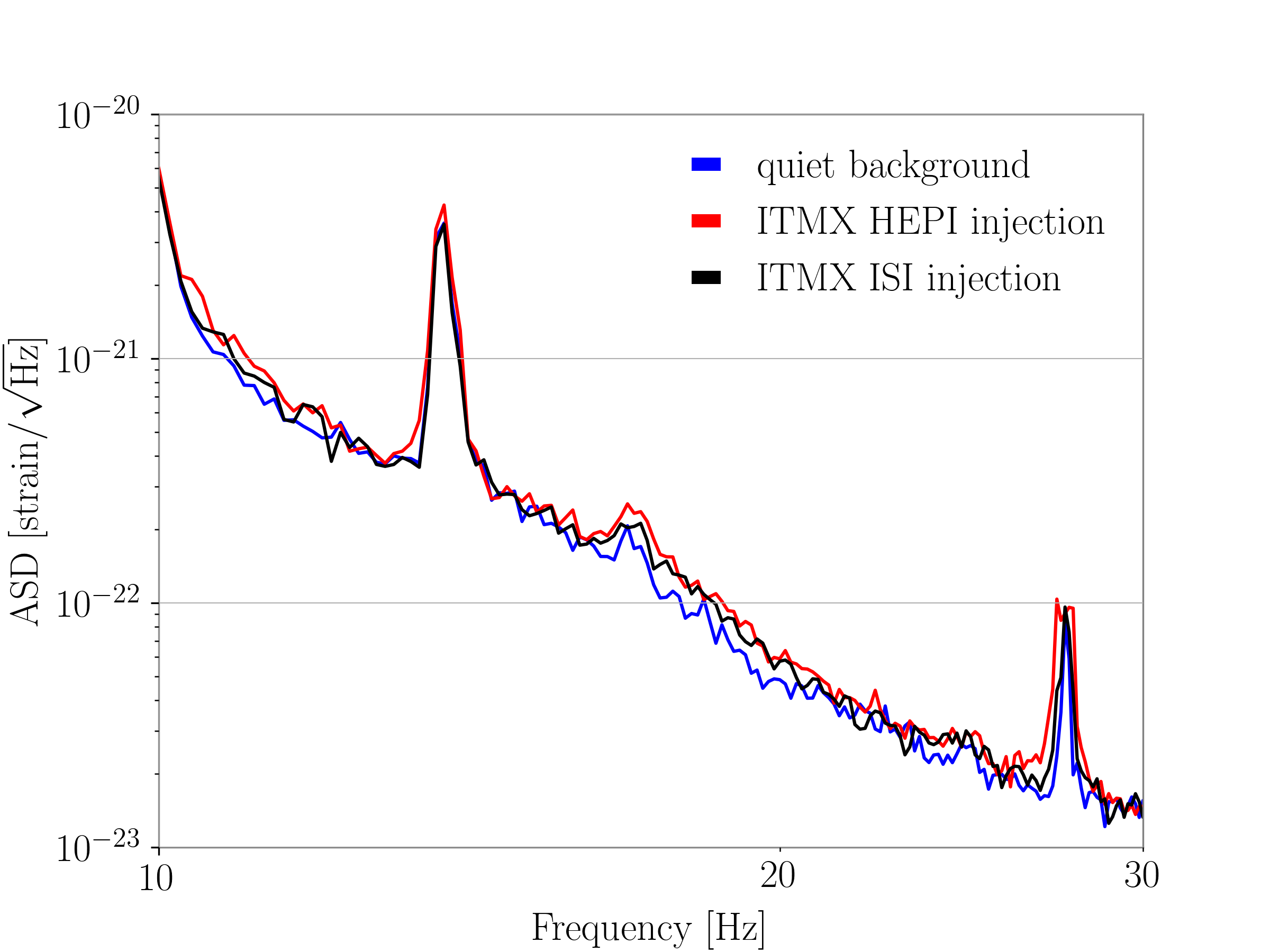}
  \caption{}
  \label{fig:itmx_injection_DARM}
\end{subfigure}
\caption{Results of injecting $0.2$ Hz motion at ITMX; (a) Velocity at ITMX ISI stage $2$ during the $0.2$ Hz injections compared with a normal background; (b) Noise produced in DARM by these injections compared with a suitable background.}
\label{fig:itmx_injection}
\end{figure*}

\begin{figure*}
\centering
\begin{subfigure}{.45\textwidth}
  \centering
  \includegraphics[width=1.0\textwidth]{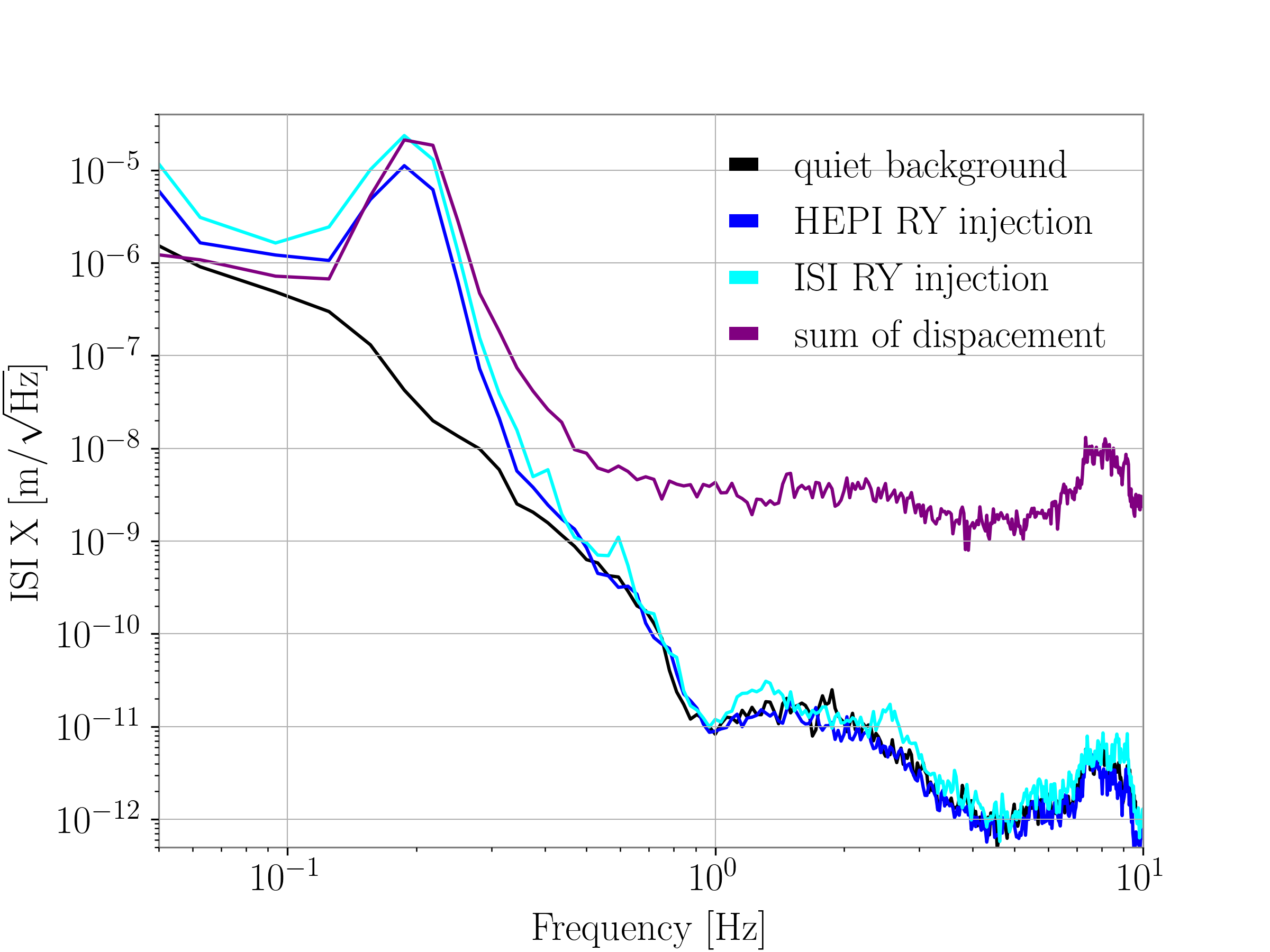}
  \caption{}
  \label{fig:bs_inj_RY}
\end{subfigure}%
\begin{subfigure}{.45\textwidth}
  \centering
  \includegraphics[width=1.0\textwidth]{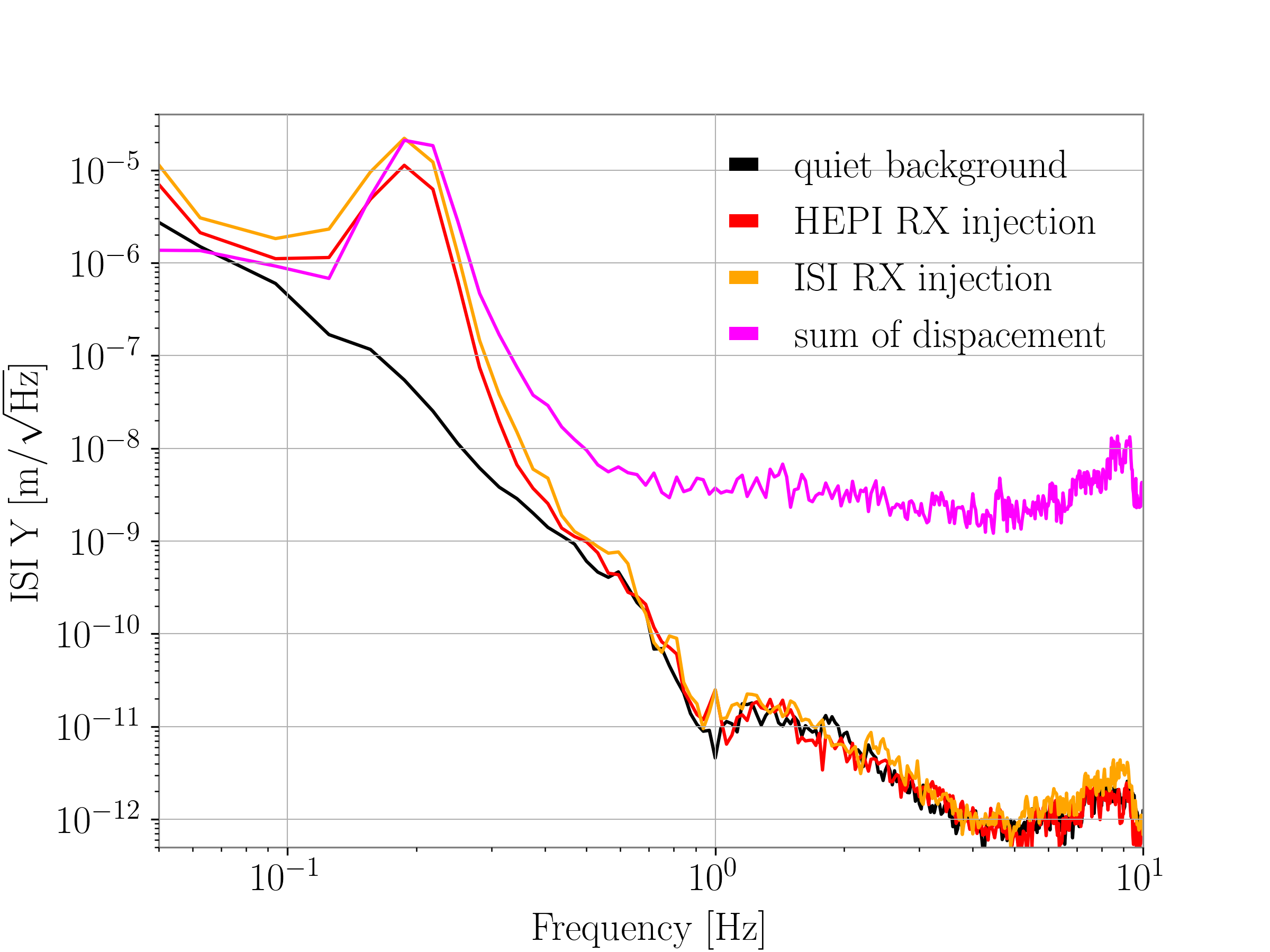}
  \caption{}
  \label{fig:bs_inj_RX}
\end{subfigure}
\caption{Translational motion at the BS ISI stage $2$ during the $0.2$ Hz injections; (a) Injections along RY. The dark blue and cyan traces show the motion as recorded by the inertial sensor (GS13) of ISI stage-2 during the HEPI and ISI RY injections respectively. The purple trace shows a sum of the displacement sensors of HEPI, ISI stage-1 and ISI stage-2. The total displacement along X direction as measured by the displacement sensors at around $0.2$ Hz are almost 3 times higher than that measured by the inertial sensor; (b) Injections along RX. The orange and red traces show the motion as recorded by the inertial sensor (GS13) of ISI stage-2 during the HEPI and ISI RX injections respectively. The magenta trace shows a sum of the displacement sensors of HEPI, ISI stage-1 and ISI stage-2. Here also, the total displacement along Y direction at around $0.2$ Hz as measured by the displacement sensors are almost 3 times higher than that measured by the inertial sensor.}
\label{fig:bs_inj_motion}
\end{figure*}

\section{Instrumental investigations}
\label{investigations}
In order to find out the source of scattered light noise in O4, we checked a few suspects which had created noise during the third observation run (O3) and proved that they are not producing noise by the same coupling mechanisms anymore. As a part of our investigations, we injected motion at microseismic frequencies at the different stages of seismic isolation platforms at all the test masses and at the beam splitter and measured the level of noise produced by them. In this section, we discuss the results of all of these injections. 

 To begin with, we injected longitudinal motion at $0.2$ Hz and $0.32$ Hz in the HEPI and ISI platforms of all test mass mirrors. When motion was injected at the HEPI, it moved the ACB and the quadruple suspension chain. We again injected the same amount of motion at the same frequencies at stage $2$ of the ISI. This moved the test mass mirror by the same amount as before but not the ACB and we did not notice any difference in the level of noise in DARM produced by these two injections. Figure \ref{fig:itmx_injection_motion} shows an example of the amount of motion that was injected at the ITMX ISI during the HEPI and ISI injections at $0.2$ Hz, and figure \ref{fig:itmx_injection_DARM} shows the noise produced in DARM by this much motion. We repeated the same injections at all the test masses and got the highest amount of signal in DARM from the injections at ITMX, though the DARM signal produced by the ITMX injections was almost $5$ times smaller in amplitude than the scatter shelf seen in DARM during high microseismic ground motion.

\begin{figure}
  \centering
  \includegraphics[width=0.45\textwidth]{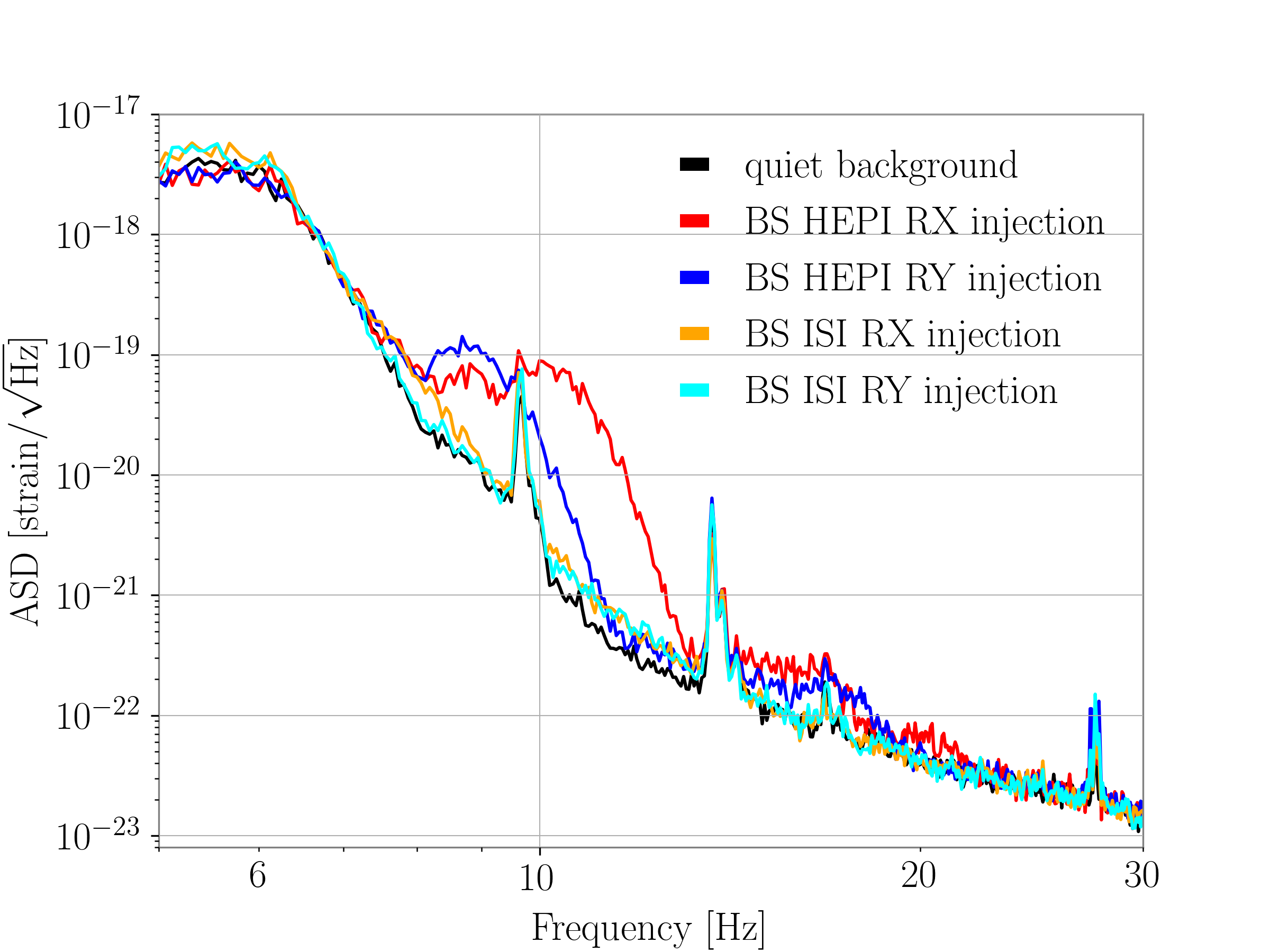}
  \caption{Noise produced in DARM by the $0.2$ Hz BS injections compared with a suitable background}
  \label{fig:bs_inj_darm}
\end{figure} 

\begin{figure*}
\centering
\begin{subfigure}{.45\textwidth}
  \centering
  \includegraphics[width=1.0\textwidth]{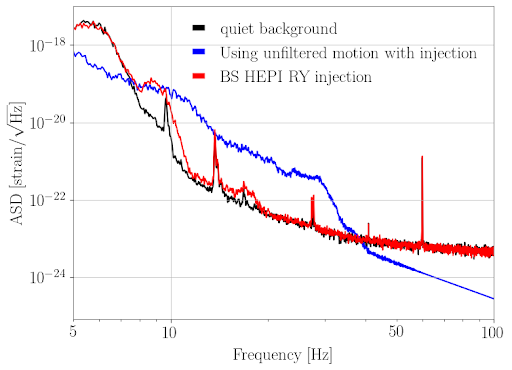}
  \caption{}
  \label{fig:bs_inj_w_raw_motion_model}
\end{subfigure}%
\begin{subfigure}{.45\textwidth}
  \centering
  \includegraphics[width=1.0\textwidth]{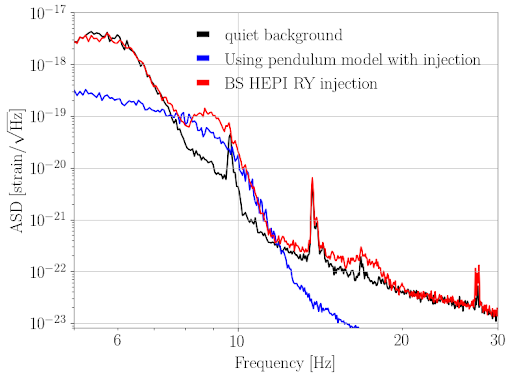}
  \caption{}
  \label{fig:bs_injections_w_filtered_motion_model}
\end{subfigure}
\caption{Scatter shelves created by combining the models with injected motion; (a) The blue trace shows the scatter shelf calculated using the injected motion without any filtering and the red trace shows the actual scatter shelf obtained while injecting motions at the BS HEPI. (b) The blue trace shows the scatter shelf calculated using the injected motion filtered by the pendulum transfer function and the red trace shows the actual scatter shelf obtained during the injections at BS HEPI.}
\label{fig:model+injections}
\end{figure*} 

Similarly, we injected $0.2$ Hz motion at the BS HEPI and ISI along different degrees of freedom (DOFs). The HEPI injections along the RY (rotation along Y) and RX (rotation along X) DOFs produced big scatter shelves as shown in figure \ref{fig:bs_inj_darm}. However, due to cross-couplings between different DOFs, RY and RX injections also produced a significant amount of longitudinal motion along X and Y, respectively, as shown in figure \ref{fig:bs_inj_motion}. It is important to note here that only the HEPI injections produced the scatter shelves and not the ISI injections, which means it is the motion of the elliptical baffles which is responsible for creating these shelves. The scatter shelves produced by the BS injections were almost $100$ times higher in amplitude and thus $10^4$ times higher in power than what we typically observed in DARM during high microseismic ground motion. This indicates that there is more scattered beam falling on the elliptical baffles than what typically produced the scattered light noise in O4. But a fraction of this light can still get reflected back by some other surface in the vicinity and produce noise in DARM. So, the elliptical baffles may be a potential source of the noise.  

Figure \ref{fig:model+injections} shows a comparison of our models with BS HEPI injections. The blue trace in \ref{fig:bs_inj_w_raw_motion_model} shows the scatter shelf calculated using the injected motion at the BS HEPI platform without doing any filtering (as described in the model of section \ref{sec:3.1}) and using a light amplitude of $4\times10^{-8}$. The red trace in the same figure shows the actual scatter shelf that we obtained while injecting the motion at the BS HEPI platform. Figure \ref{fig:bs_injections_w_filtered_motion_model} shows a similar comparison with the other model in which we filtered the motion of the HEPI platform by the pendulum transfer function, as described in section \ref{sec:3.2}. The blue trace in figure \ref{fig:bs_injections_w_filtered_motion_model} shows the scatter shelf calculated using the injected motion and using a light amplitude of $4\times10^{-9}$. It is important to note here that for both models, we had to use light amplitudes of one order of magnitude higher (this means $100$ time more in power) to fit the scatter shelves produced by BS injections than what we typically use to fit the scatter shelves of O4 noise. This again indicates that there is much more scattered light falling on the elliptical baffles than what typically produced the noise in O4. In summary, both of these models can explain possible noise coupling mechanisms but they require different light amplitudes to fit the scatter shelves because they explain two different scatter paths. For the injections, the model with filtered motion fits better with the scatter shelf and this can be due to the fact that by injecting motion along a specific degree of freedom, we are mimicking one of the scatter paths better than the other.  

\section{20 Hz glitches}
\label{sec:20_Hz_glitches}
\begin{figure*}
\centering
\begin{subfigure}{.5\textwidth}
  \centering
  \includegraphics[width=1.1\textwidth]{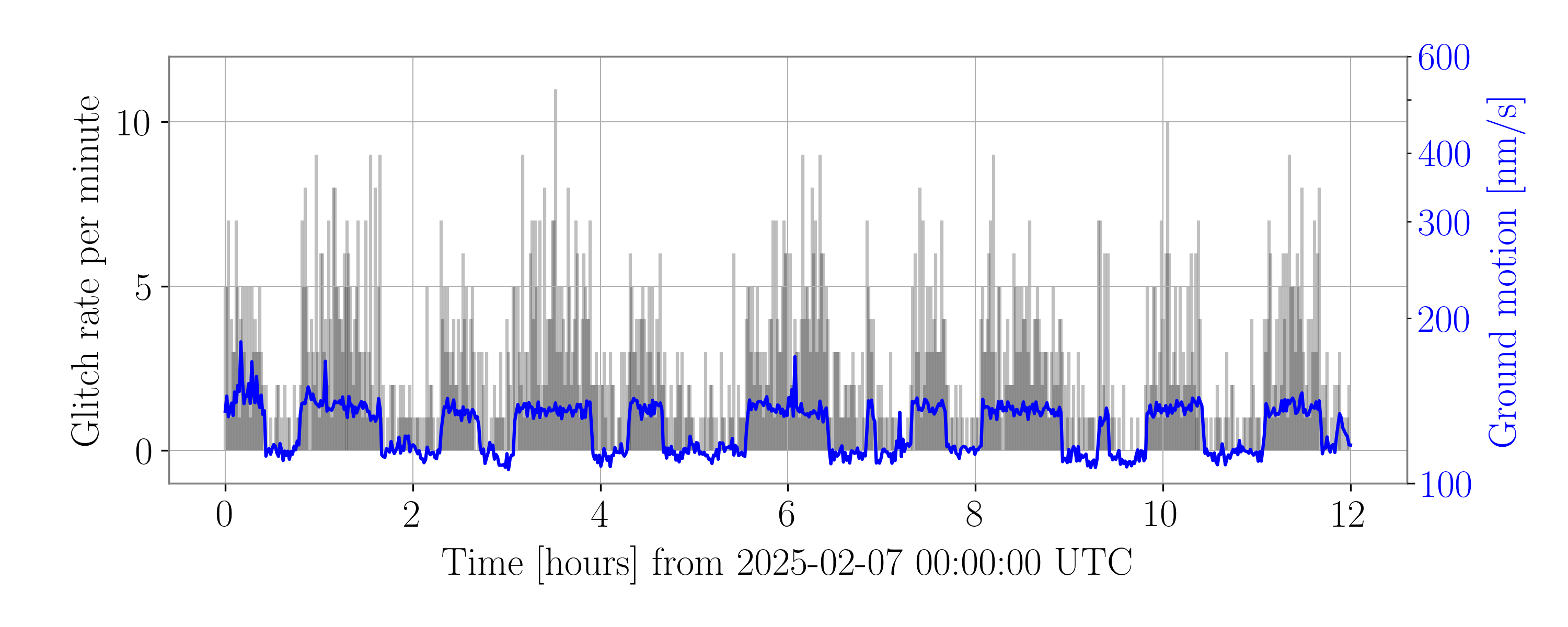}
  \caption{}
  \label{fig:20_Hz_glitch_vs_gm_pre}
\end{subfigure}%
\begin{subfigure}{.4\textwidth}
  \centering
  \includegraphics[width=0.8\textwidth]{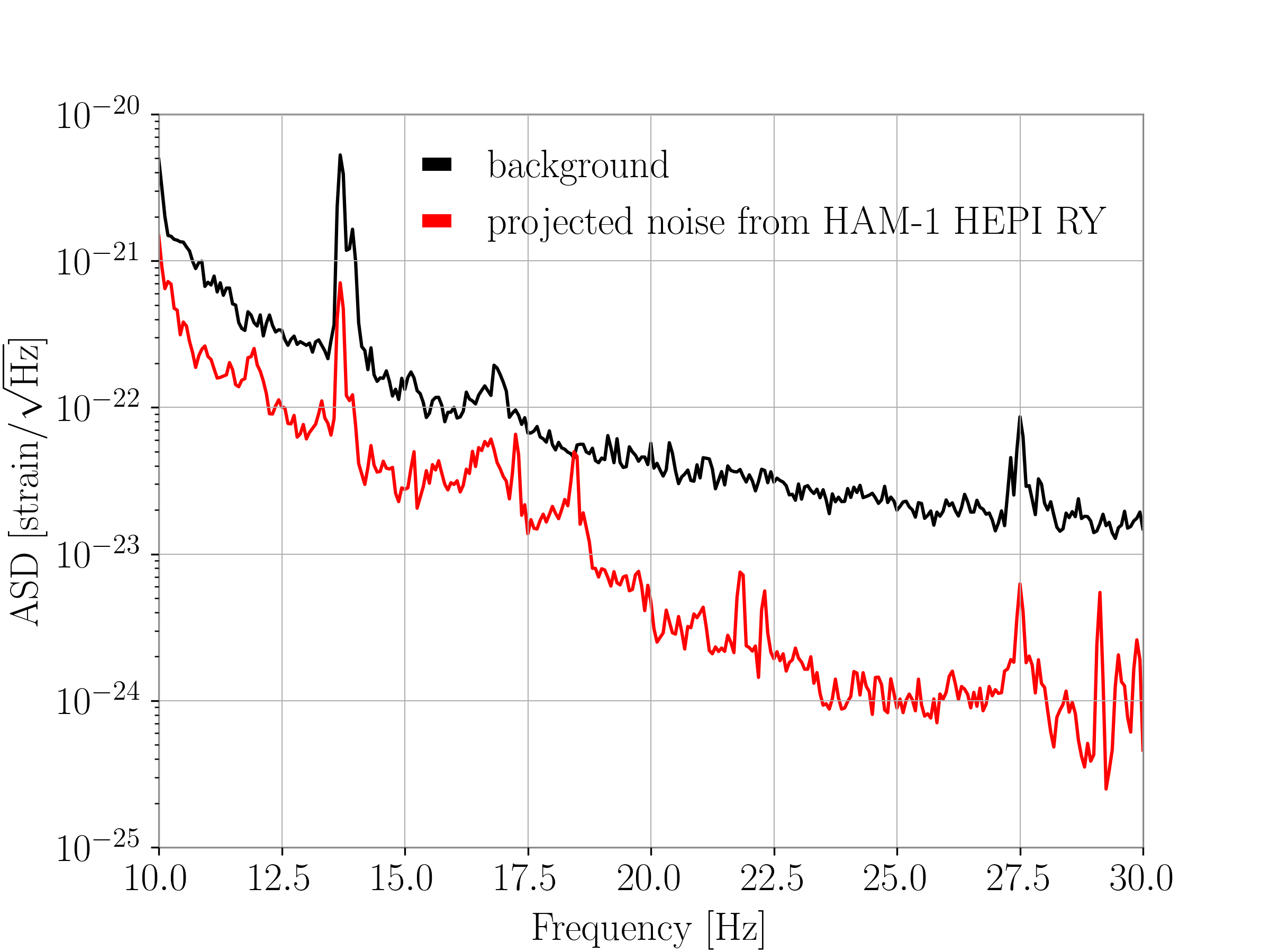}
  \caption{}
  \label{fig:HAM1_noise_projection_pre}
\end{subfigure}
\caption{(a) Overlaid plot of glitch rate per minute and the vertical ground motion at the corner station in the $10$-$30$ Hz frequency band (before installing the ISI); (b) Projected noise from the HAM-1 HEPI motion in RY DOF.}
\end{figure*}

Another group of glitches appeared in the $15$-$25$ Hz frequency band with lower SNR compared to the glitches we have discussed so far. In the omicron diagram of figure \ref{fig:omicron}, they are visible as a group of low SNR glitches (light green band) around $20$ Hz. We have learned that the rate of these glitches is well correlated with the vertical ground motion at the corner station in the $10$-$30$ Hz frequency band. Figure \ref{fig:20_Hz_glitch_vs_gm_pre} shows an overlaid plot of vertical ground motion in the $10$-$30$ Hz frequency band at the corner station (in blue trace) and the rate of the low SNR glitches (in gray). Even though these glitches are primarily modulated by the high frequency ground motion, they also have some correlations with the microseismic ground motion, appearing only when the microseismic ground motion is above a certain threshold along with the $10$-$30$ Hz ground motion.   

\begin{figure*}
\centering
\begin{subfigure}{.5\textwidth}
  \centering
  \includegraphics[width=1.1\textwidth]{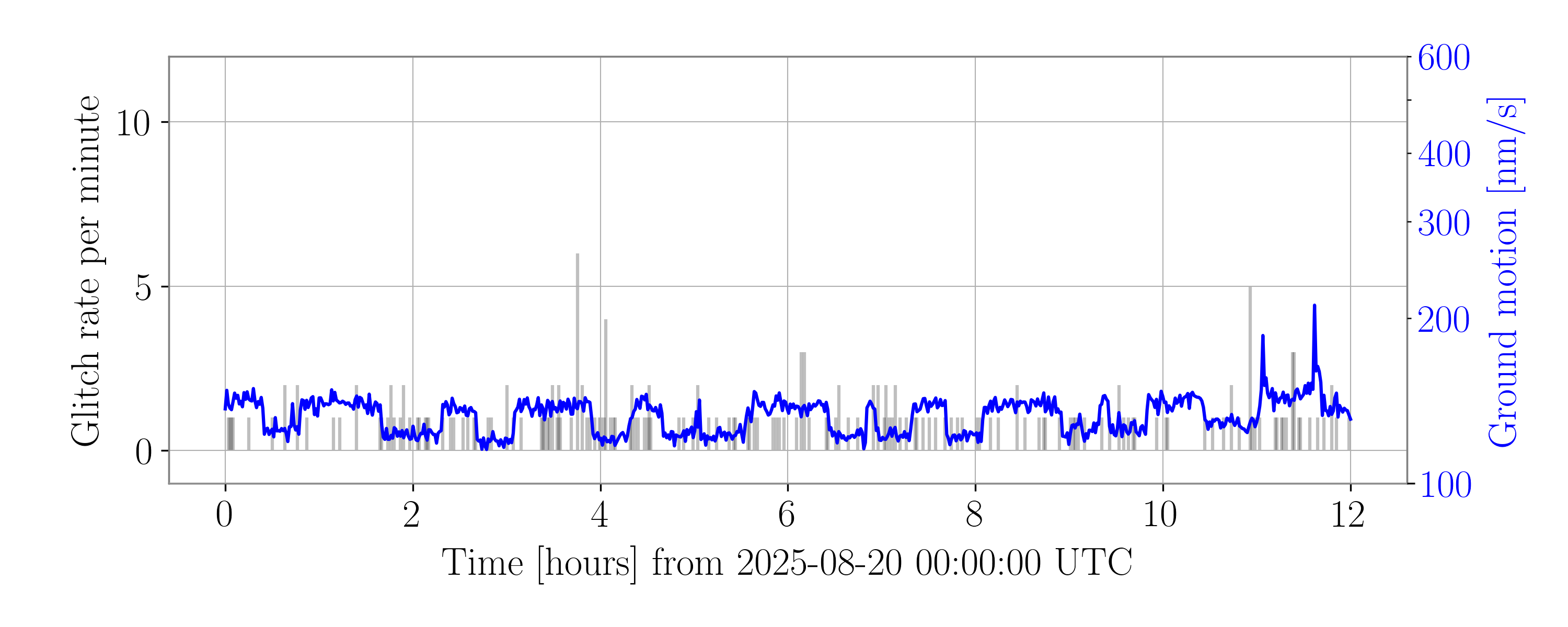}
  \caption{}
  \label{fig:20_Hz_glitch_vs_gm_post}
\end{subfigure}%
\begin{subfigure}{.4\textwidth}
  \centering
  \includegraphics[width=0.8\textwidth]{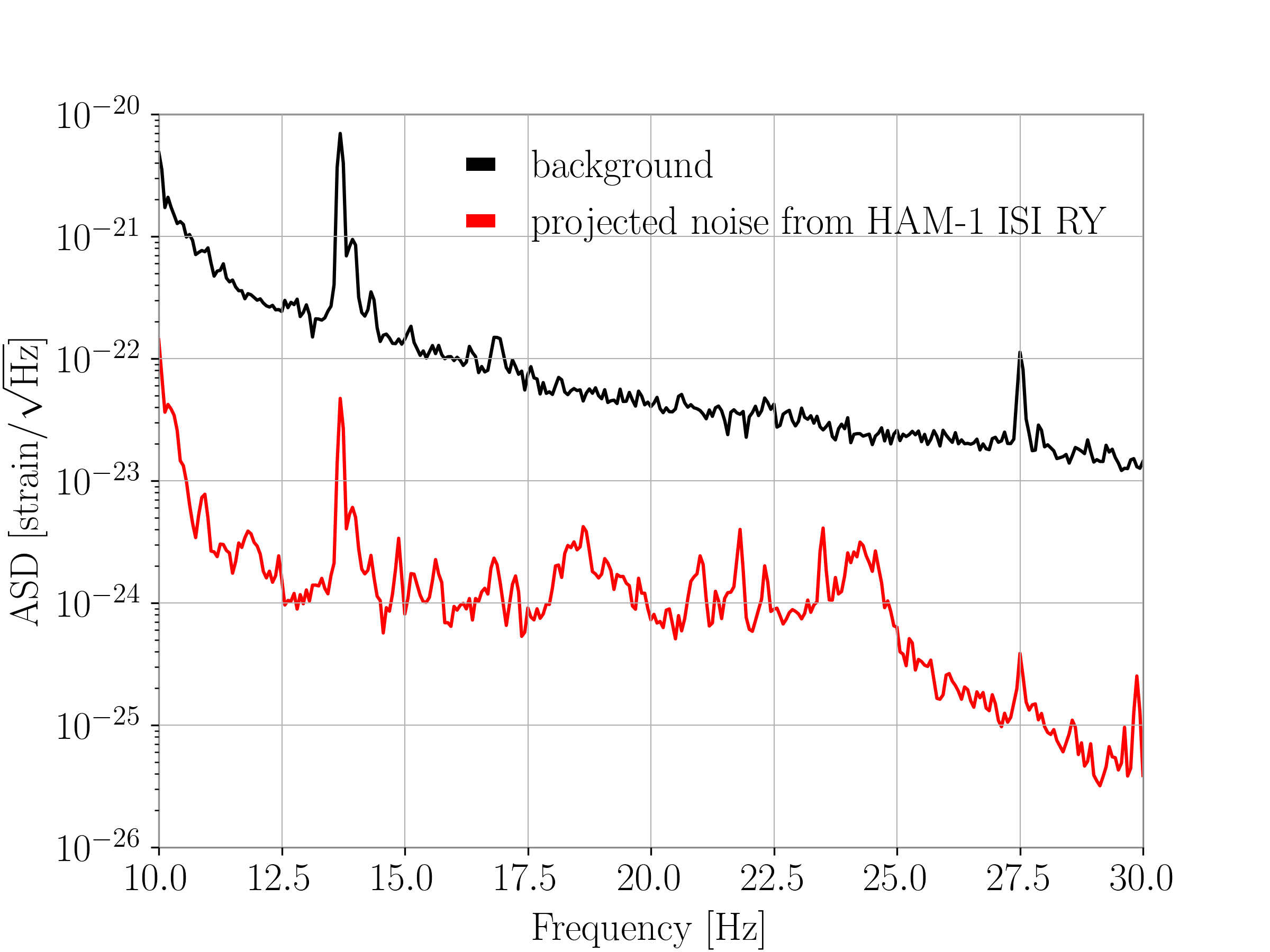}
  \caption{}
  \label{fig:HAM1_noise_projection_post}
\end{subfigure}
\caption{(a) Overlaid plot of glitch rate per minute and the vertical ground motion at the corner station in the $10$-$30$ Hz frequency band (after installing the ISI); (b) Projected noise from the HAM-1 ISI motion in RY DOF. } 
\end{figure*}

We suspected that high frequency ground motion was coupling through the Horizontal Axis Module-1 (HAM-1) chamber (the vacuum chamber next to the pre-stabilized laser system) at the corner station after measurements revealed that the rotational motion of the HEPI platform at the HAM-1 chamber is strongly coupled to DARM. We used these measurements of coupling functions to calculate the projected noise from the motion of the HEPI platform. Figure \ref{fig:HAM1_noise_projection_pre} shows the projected noise from the motion of the HEPI platform to be very close to the noise spectra of LLO. Also, HAM-1 was the only vacuum chamber in the corner station without an ISI platform. During the last vent (summer $2025$) at LLO, an ISI platform was installed in this chamber and since then, these glitches have not appeared in the data. Figure \ref{fig:20_Hz_glitch_vs_gm_post} shows an overlaid plot of glitch rate with the $10$-$30$ Hz ground motion after installing the ISI platform and, when compared to the glitch rate in figure \ref{fig:20_Hz_glitch_vs_gm_pre}, it is clear that the glitch rate is significantly reduced and there is no longer any correlation between the glitch rate and the ground motion. Figure \ref{fig:HAM1_noise_projection_post} shows the projected noise from the motion of the ISI platform; it is much below the noise spectra of LLO and hence the previously suspected coupling through the HAM-1 chamber is no longer there. Though, it is not clear exactly which surface and which beam inside the HAM-1 chamber were responsible for this noise. 

\section{Impact of instrumental changes at LLO} 
\label{sec:instrumental_upgrades}

\begin{figure*}
\centering
\begin{subfigure}{.5\textwidth}
  \centering
  \includegraphics[width=0.9\textwidth]{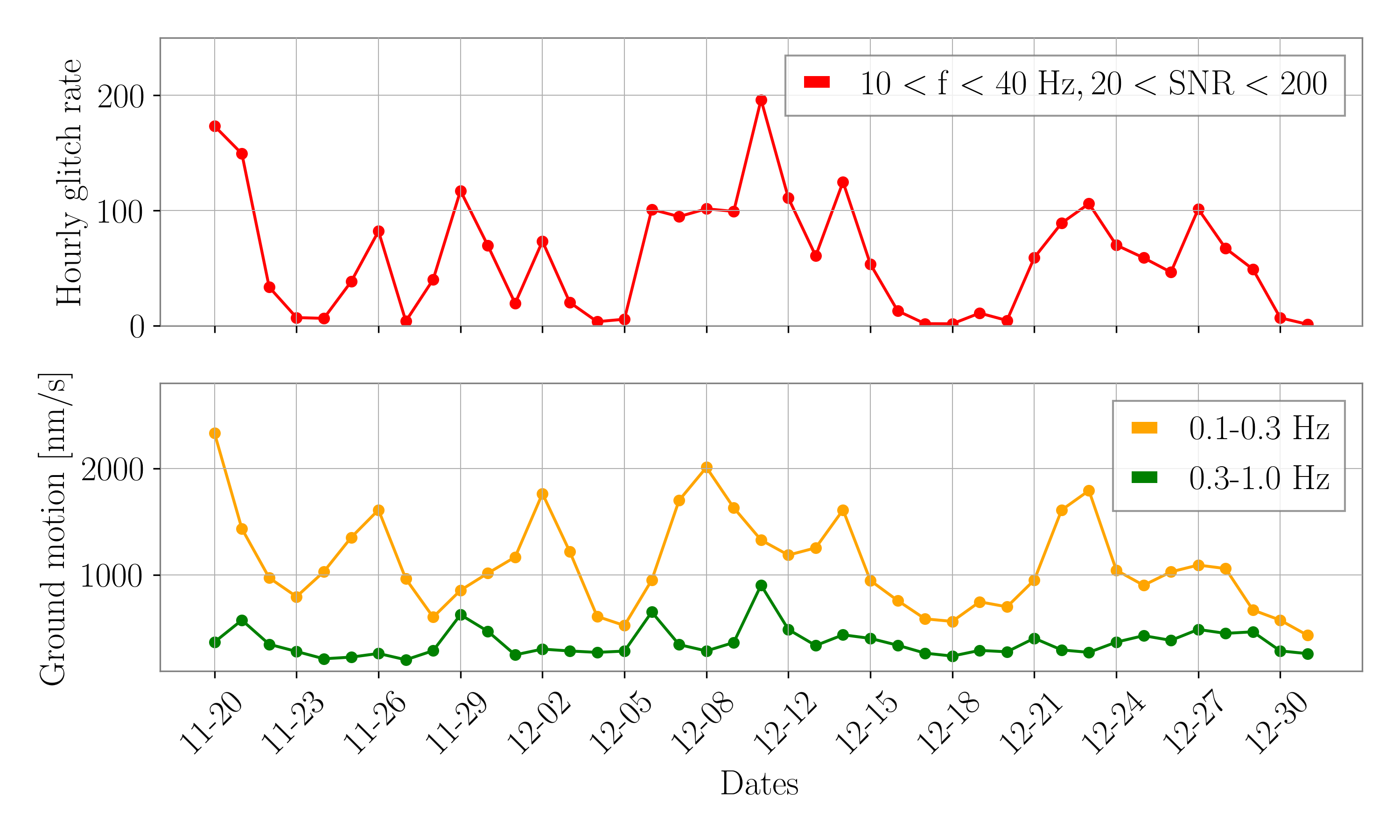}
  \caption{}
  \label{fig:high_snr_vs_gm_pre}
\end{subfigure}%
\begin{subfigure}{.5\textwidth}
  \centering
  \includegraphics[width=0.9\textwidth]{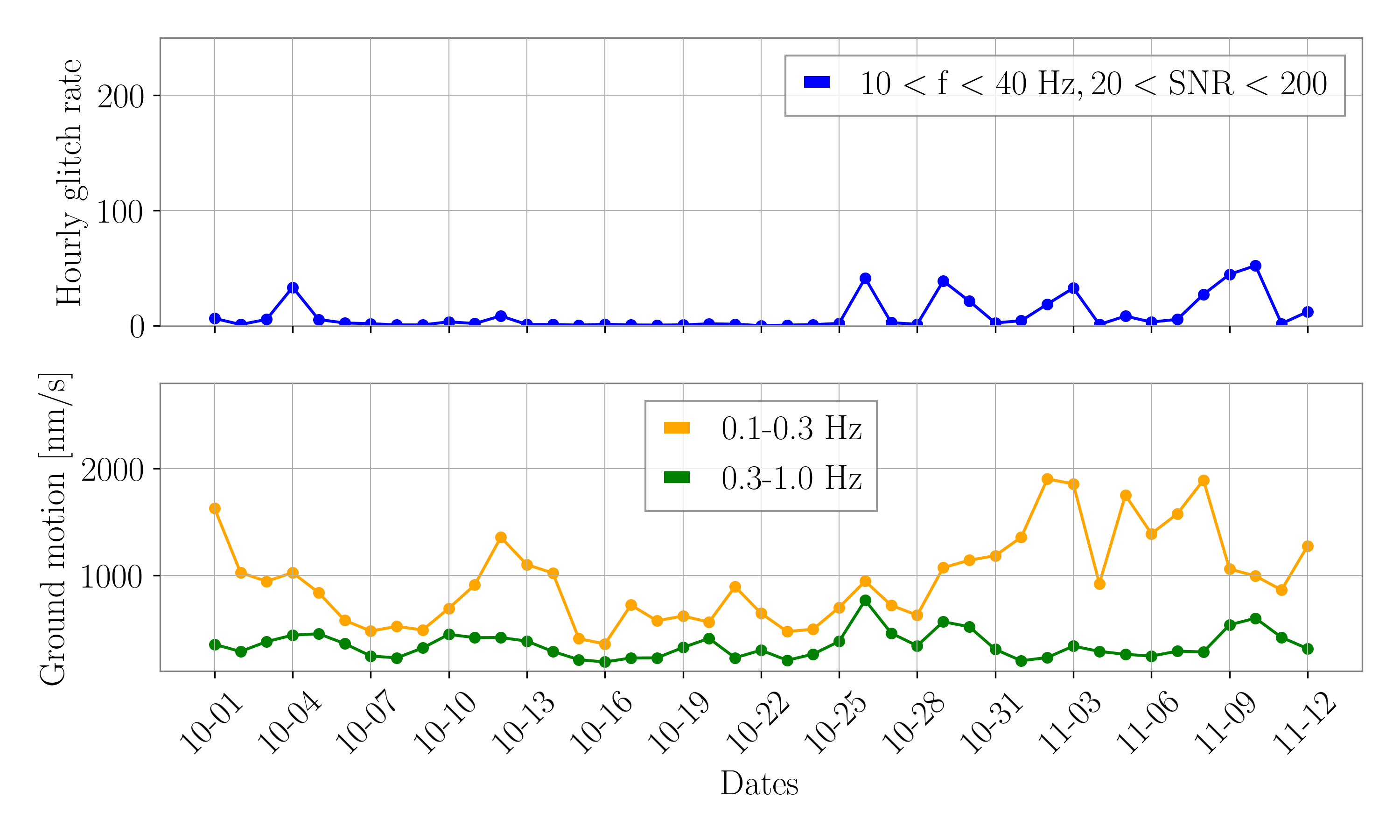}
  \caption{}
  \label{fig:high_snr_vs_gm_post}
\end{subfigure}
\caption{(a) \textit{Top}: Hourly rate of glitches from $20$th November $2024$ to $31$st December $2024$ (before the O4C vent); glitches chosen for this analysis have frequency in $10$-$40$ Hz and SNR in $20$-$200$. \textit{Bottom}: Ground motion during the same period of time in two different microseismic frequency bands $0.1$-$0.3$ Hz (shown in orange trace) and $0.3$-$1.0$ Hz (shown in green trace) (b) \textit{Top}: Hourly rate of glitches from $1$st October $2025$ to $12$th November $2025$ (after the O4C vent); glitches chosen for this analysis have frequency in $10$-$40$ Hz and SNR in $20$-$200$. \textit{Bottom}: Ground motion during the same period of time in two different microseismic frequency bands $0.1$-$0.3$ Hz (shown in orange trace) and $0.3$-$1.0$ Hz (shown in green trace). The glitch rate has significantly reduced after the instrumental upgrades performed during the O4C vent.} 
\end{figure*}

During the most recent vent at LLO (April–July 2025), several key hardware upgrades were implemented to mitigate scattered light noise and improve detector stability. One of these upgrades was installing additional baffles very close to all the test mass mirrors to prevent some of the light scattering through the edge of the mirrors from re-entering the main beam path. In addition, an ISI platform was installed in the HAM-1 chamber, which provides enhanced seismic isolation to the optics table hosted therein. Collectively, these upgrades have contributed to a noticeable reduction in the glitch rate on days with high microseismic ground motion. Although some glitches are still present, their SNRs are lower than those observed prior to the vent under comparable ground motion conditions.

The top panel in figure \ref{fig:high_snr_vs_gm_pre} shows the hourly glitch rate for almost a month ($20$th November $2024$ - $31$st December $2024$) before doing the instrumental upgrades. The bottom panel of the same figure shows the microseismic ground motion in two different frequency bands ($0.1$-$0.3$ Hz and $0.3$-$1.0$ Hz) during the same period of time. Similarly, figure \ref{fig:high_snr_vs_gm_post} shows the glitch rate and ground motion for almost a month ($1$st October $2025$ - $12$th November $2025$) after completing the instrumental upgrades. Comparing the top panels of the figures \ref{fig:high_snr_vs_gm_pre} and \ref{fig:high_snr_vs_gm_post}, we can see a significant reduction in the hourly glitch rate given the same amount of microseismic ground motion. 

\begin{figure*}
\centering
\begin{subfigure}{.5\textwidth}
  \centering
  \includegraphics[width=1.0\textwidth]{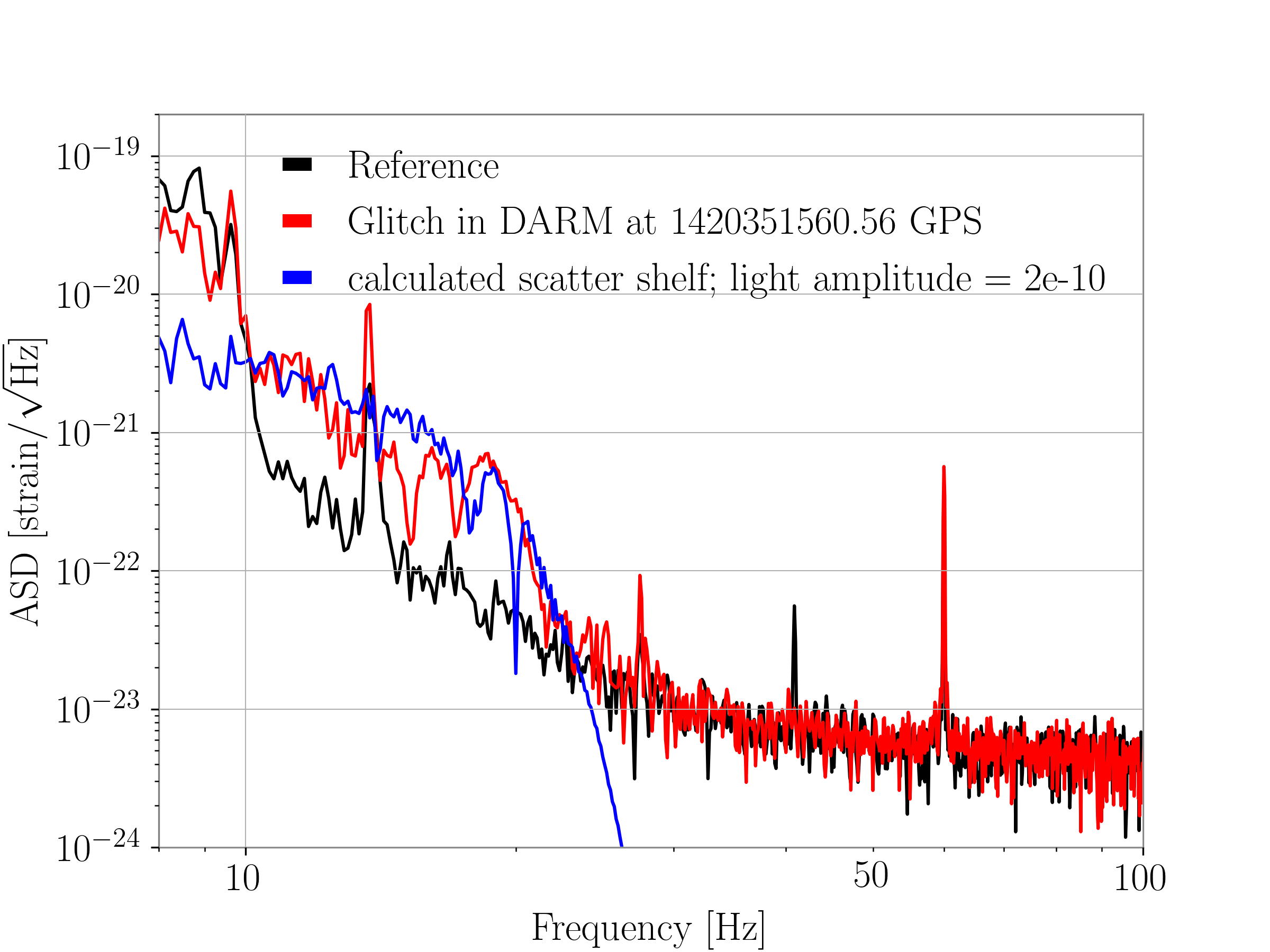}
  \caption{}
  \label{fig:light_amp_pre_vent}
\end{subfigure}%
\begin{subfigure}{.5\textwidth}
  \centering
  \includegraphics[width=1.0\textwidth]{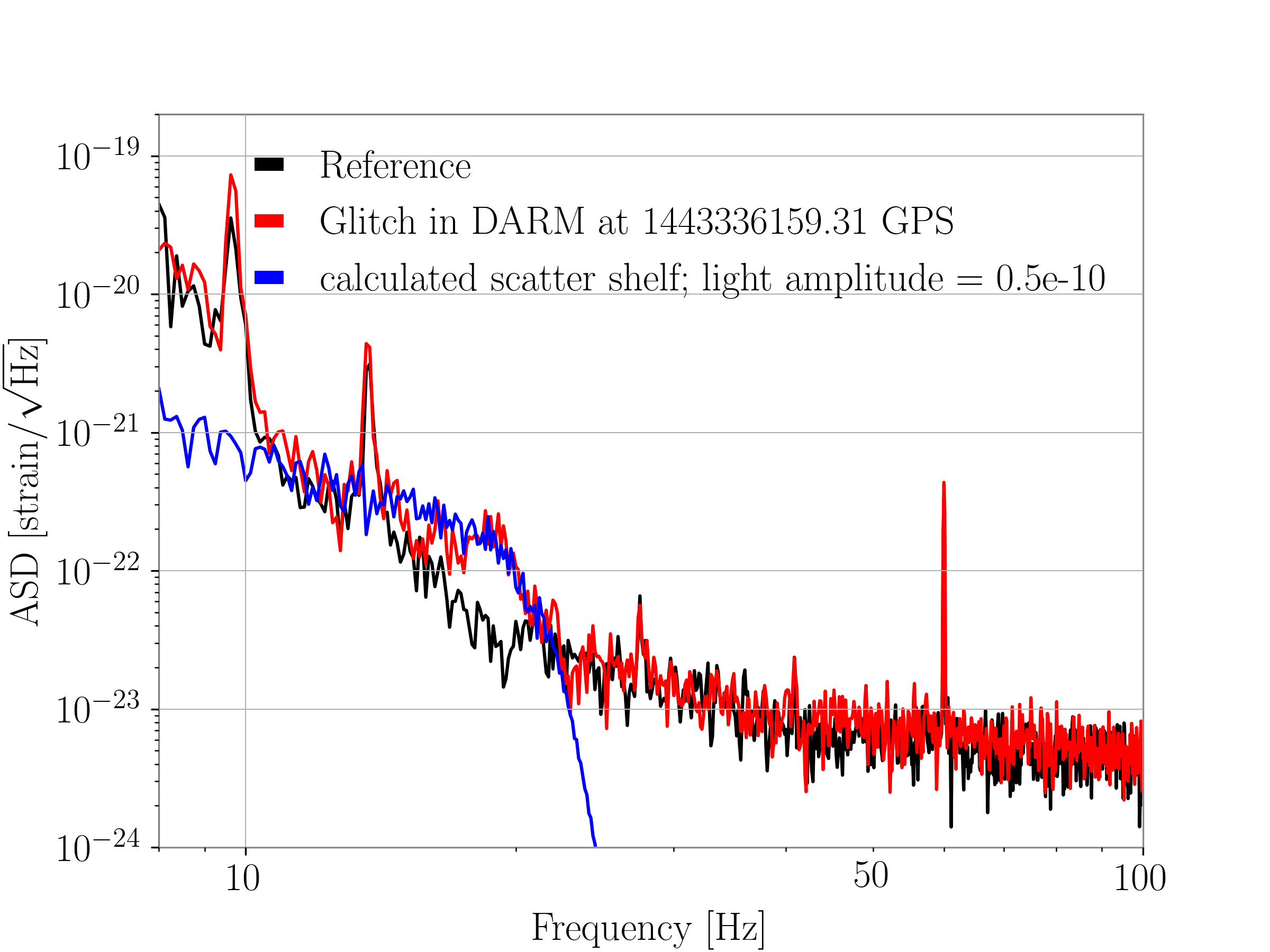}
  \caption{}
  \label{fig:light_amp_post_vent}
\end{subfigure}
\caption{Comparing light amplitude before and after the cage baffle installation; (a): Scatter shelf before installing the additional baffles; (b): Scatter shelf after installing the additional baffles.}
\label{fig:light_amp_comparison}
\end{figure*}

\begin{figure}
    \centering
    \includegraphics[width=0.5\textwidth]{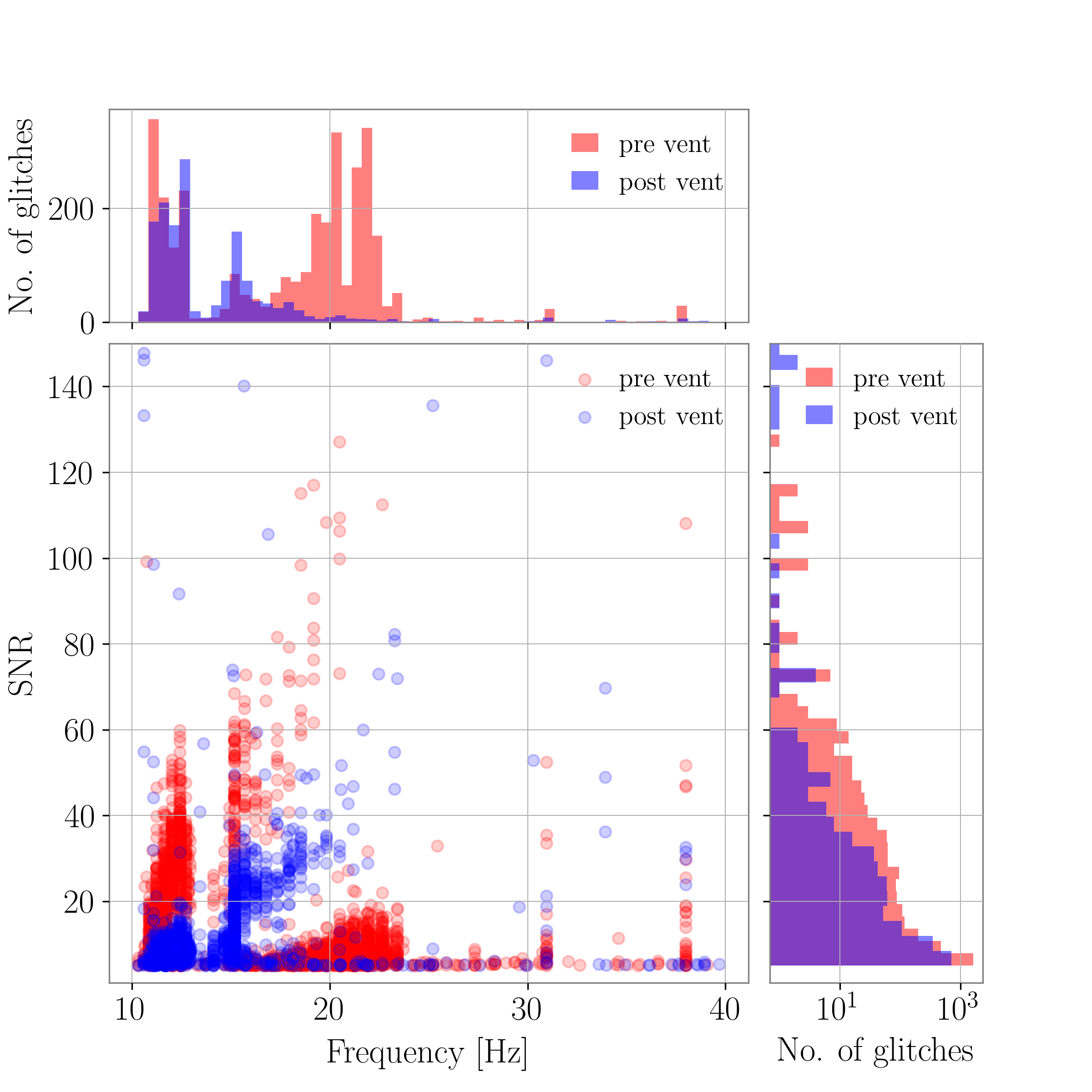}
    \caption{Frequency and SNR distribution of the glitches during two days, one before doing the instrumental upgrades (2025-01-08 16:00:00-23:59:59 UTC) and another after the changes (2025-11-08 16:00:00-23:59:59 UTC). These two days had very similar microseismic ground motion. The red dots in the scatter plot show the frequency and SNR of the glitches before the instrumental changes were made and the blue ones show the same after the changes. The histograms in the right side of the scatter plot show the distribution of SNR of the glitches, clearly showing the reduction in SNR after performing the instrumental upgrades. The histograms on the top of the scatter plot show the frequency distribution of the glitches before and after the instrumental changes; it clearly indicates a reduction in the number of glitches, especially around 20 Hz.}
    \label{fig:snr_freq_comparison}
\end{figure}

Figure \ref{fig:light_amp_comparison} compares two scatter shelves, one from before the baffles were installed (figure \ref{fig:light_amp_pre_vent}) and the other from after baffle installation (figure \ref{fig:light_amp_post_vent}). Even though both of these days had very similar microseismic ground motion, it is evident that the height of the shelf reduced after installing the baffles. The blue traces in each plot of figure \ref{fig:light_amp_comparison} show the scatter shelf calculated using filtered motion (model described in section \ref{sec:3.2}). For the figure \ref{fig:light_amp_pre_vent} (one without the baffles), we used a light amplitude of $2\times10^{-10}$ to match the actual scatter shelf and for the figure \ref{fig:light_amp_post_vent} (one with the baffles), the required light amplitude to match the actual scatter shelf was $0.5\times10^{-10}$. 

Figure \ref{fig:snr_freq_comparison} shows the SNR and frequency distributions of glitches occurring on two different days that had very similar microseismic ground motion, with one of the two days being before the instrument upgrades were completed and the other, after the additional baffles and the ISI platform in the HAM-1 vacuum chamber were installed. The histogram on top of the scatter plot shows the frequency distributions of the glitches on these two days and there is a significant reduction in the number of glitches, especially around 20 Hz. The histograms on the right side of the scatter plot show the SNR distributions of the glitches of these two days, and a clear reduction in the SNR of the glitches occurring after performing the instrumental upgrades. From this plot, it is evident that after installing the additional baffles and the seismic isolation platform (ISI) in the HAM-1 vacuum chamber, not only did the glitch rate significantly drop but also, the SNRs of the remaining glitches were reduced. 

\section{Scenario at LHO} 
\label{sec:LHO}

\begin{figure}
    \centering
    \includegraphics[width=0.45\textwidth]{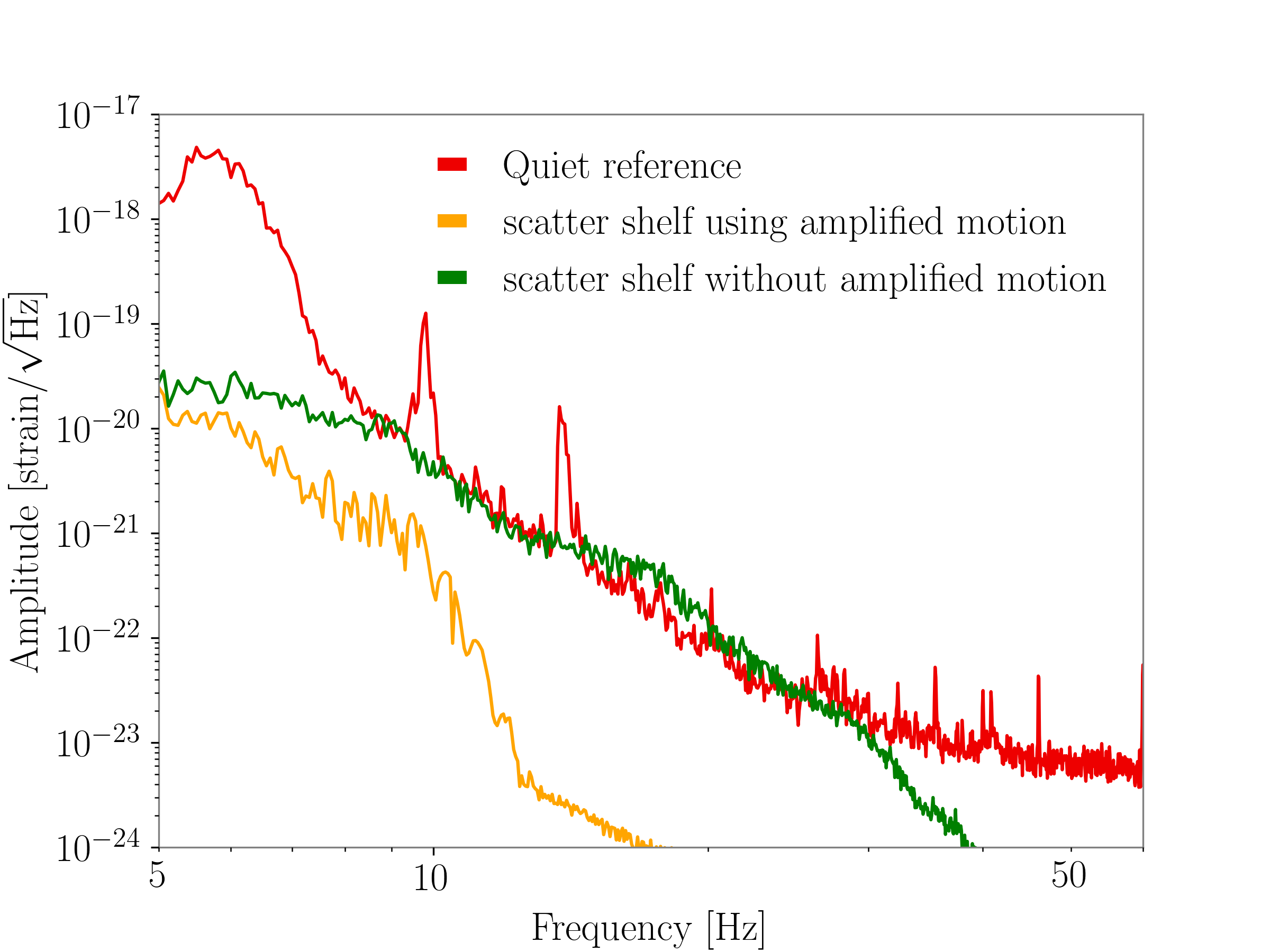}
    \caption{Simulated scatter shelf and the current noise spectra of the LIGO Hanford detector.}
    \label{fig:LHO_shelf}
\end{figure}

Since both LIGO observatories have a very similar design and operating mechanism, it is possible that some noise arising in one of the detectors might appear in the other detector too if it is due to an artifact of the instrumental design. So we checked whether a scattering mechanism similar to LLO can produce noise at LHO. We filtered the microseismic ground motion at LHO using the modeled transfer function of a potential scattering surface, as described in section \ref{sec:3.2} and assumed the amount of light in the scattered beam to be the same as LLO. The simulated noise spectra, as given by equation \ref{eq:scatter_eq}, is shown by the orange trace in figure \ref{fig:LHO_shelf} and it is below the present noise spectra of LHO. We then calculated the scatter shelf using the unfiltered motion (model described in section \ref{sec:3.1}) as well. The simulated noise spectra is shown by the green trace in figure \ref{fig:LHO_shelf} and it is slightly above the present DARM sensitivity of LHO, especially around 20 Hz. But this is assuming the same amount of light as LLO. If the amount of scattered light in case of LHO is slightly less than that of LLO, then this scatter shelf would still remain below the noise spectra of LHO. This is primarily because of the fact that the microseismic ground motion is much lower at LHO as compared to that of LLO.

\section{Conclusions and future work}
\label{conclusions}
In summary, our analysis of the O4 data has identified two distinct groups of glitches in the data of LLO likely created by scattered light: a low SNR and high SNR group. The low SNR group was linked to a combination of anthropogenic motion and microseism motion and was abated after the installation of more seismic isolation in the vacuum chamber identified in our correlations. The high SNR group has been simulated with two models: one, directly using the motion of various surfaces, and one further applying a pendulum transfer function to the input motion of the scattered light mechanism. Both models are possible avenues of noise coupling, and it depends on the characteristics of the source which one would be correct. 

Following the installation of baffles very close to the test mass mirrors, we observed a clear reduction in both the SNR and hourly rate of the high SNR glitch group throughout most of the observation period. However, we also noticed that the glitch rate reduction depended on the frequency of the microseism, and there, there was much less reduction for glitches caused by the higher frequency microseism ($>0.3$ Hz). One possibility is that we have multiple sources of such scatter mechanisms and we have only abated one with the aforementioned baffle installation. Future work will focus on further characterizing this correlation, as well as studying the remaining glitches.

We performed a statistical correlation analysis based on these two models and both the models consistently indicated that the microseismic ground motion at the corner station along the X-arm direction has the strongest correlation with the noise created by the high SNR glitches. We also performed instrumental injections mimicking microseismic motion to try to locate the scattering source. The most promising source found was the elliptical baffles hanging near the beam splitter optic, but the signal in DARM did not match exactly the glitch parameters we were investigating. However, we can only actively test some specific coupling mechanisms and not others, so the presence of sufficient scattered light in the area could indicate that this is a possible source but with a different coupling mechanism (e.g., a moving surface we cannot excite ourselves). 

We have also verified that any coupling mechanism similar to LLO cannot produce noise in LHO primarily because the amplitude of microseismic ground motion at LHO is much less compared to that of LLO. 

In addition, the noise modeling methods explained in this paper can be used in the future to calculate projected levels of scattered light noise between various different surfaces. The correlation analysis that we implemented for localizing the sources of noise in O4 can also be implemented as a generalized method to localize sources of noise in future observation runs.

\begin{acknowledgments}
This material is based upon work supported by NSF's LIGO Laboratory which is a major facility fully funded by the National Science Foundation. LIGO was constructed
by the California Institute of Technology and Massachusetts Institute of Technology with funding from the National Science Foundation, and operates under Cooperative Agreement PHY-1764464. Advanced LIGO was built under grant No. PHY-0823459. This work uses the LIGO computing clusters and data from the Advanced LIGO detectors at Livingston, Louisiana and Hanford, Washington. The authors would like to acknowledge the support from the NSF grant PHYS-2409740. We also acknowledge the discussions with the members of the Detector Characterization working group of the LIGO Scientific Collaboration. We are thankful to Dr. Brian Lantz for providing very helpful comments and suggestions. 
\end{acknowledgments}

\appendix

\section{OSEMs at the PUM stage of ETMs}
\begin{figure}
    \centering
    \includegraphics[width=0.45\textwidth]{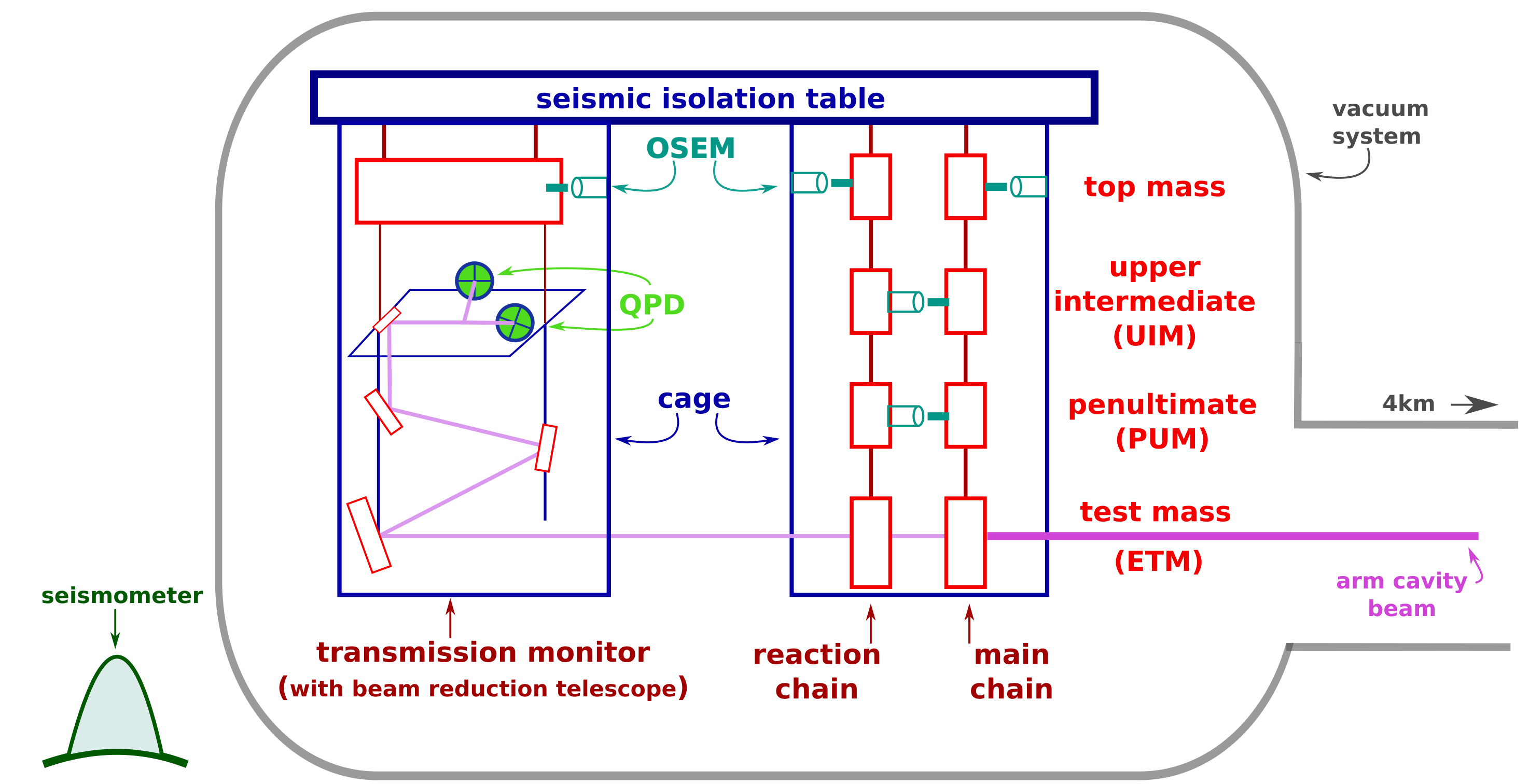}
    \caption{Schematic of the end station vacuum chamber showing the seismic isolation system, quadruple suspensions including the main and the reaction chain, and the transmitted light monitor system (TMS) \cite{LIGO:2020zwl}. The Optical Shadow sensor and Magnetic actuators (OSEMs) measure the relative motion either with respect to the suspension cage or between the main chain and the reaction chain of the quadruple suspension. These OSEMs are also used to apply a force on the main and the reaction chain.}
    \label{fig:ETM_chamber}
\end{figure}

The Optical Sensors/ElectroMagnetic actuators are used to measure and control the relative motion of the different stages of the quadruple pendulum, either with respect to the suspension cage (for the top most stage) or with respect to the reaction chain (for the UIM and the PUM stages (see figure \ref{fig:ETM_chamber})) \cite{Aston:2012ona}. During O3, the light was scattered by the gold plated electrodes, which are attached to the last stage (test mass) of the reaction chain. The OSEMs attached to the penultimate stage (PUM) of the quadruple suspension, measured a high amount of motion between the main chain and the reaction chain. This high amount of relative velocity introduced a non-linear phase coupling to the laser beam, which produced the scattered light noise.

\begin{figure*}
    \centering
    \includegraphics[width=1.0\textwidth]{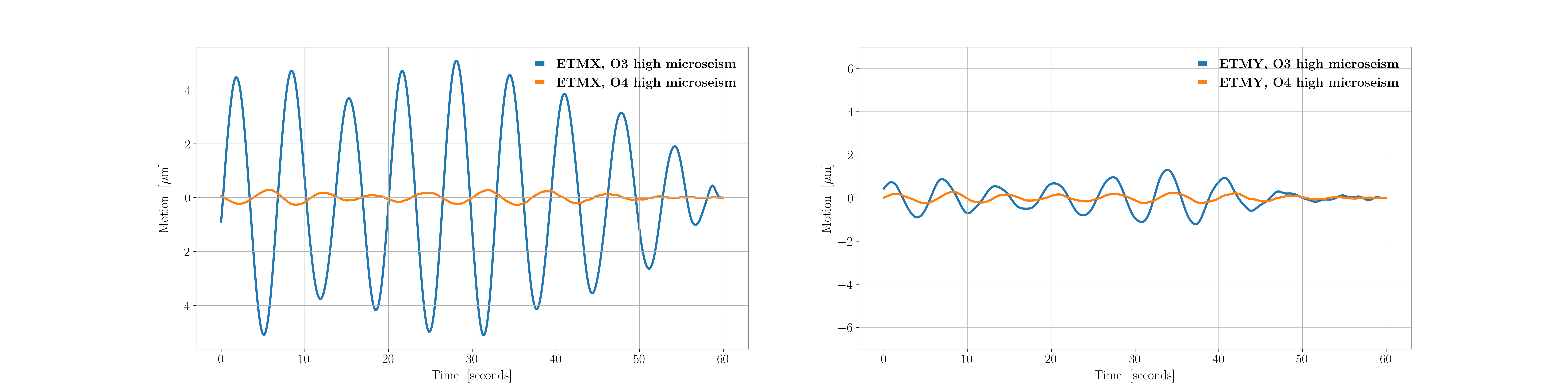}
    \caption{Motion measured by the OSEMs at the PUM stage of ETMX (\textit{Left}) and ETMY (\textit{Right}). In O3, the motion at ETMY was much less than ETMX because the control drive was applied only at the ETMX. In O4, the OSEMs measured similar amounts of motion in both the ETMs because the drive was split between the two. Also, the motion in O4 was less than that of O3 partly because of the less drive at individual ETMs and partly because of the RC tracking, which significantly reduced the motion between the main and the reaction chain of the test mass.}
    \label{fig:PUM_OSEM}
\end{figure*}

In O4, we observed that the glitches caused by scattered light had peak frequencies between $10$-$30$ Hz. This means, the scattering surface had to have a velocity of at least $5$ $\mu$m/s (calculated using equation \ref{eq:freq_prediction}) if we assume that the scattered beam was reflected only once before joining the main beam. But, unlike O3, the OSEMs of the penultimate stage did not record this much velocity in O4. 

Figure \ref{fig:PUM_OSEM} shows the time series of motion as recorded by the PUM stage OSEMs of both ETMs on a day having high microseismic ground motion. The peak-to-peak motion was around $0.4$ $\mu$m, which was much less than what is required to produce noise above $10$ Hz. This ruled out the possibility of observing any noise produced by the same coupling mechanism as O3.

\section{Scattered light in the transmitted monitors}

\begin{figure}
    \centering
    \includegraphics[width=0.5\textwidth]{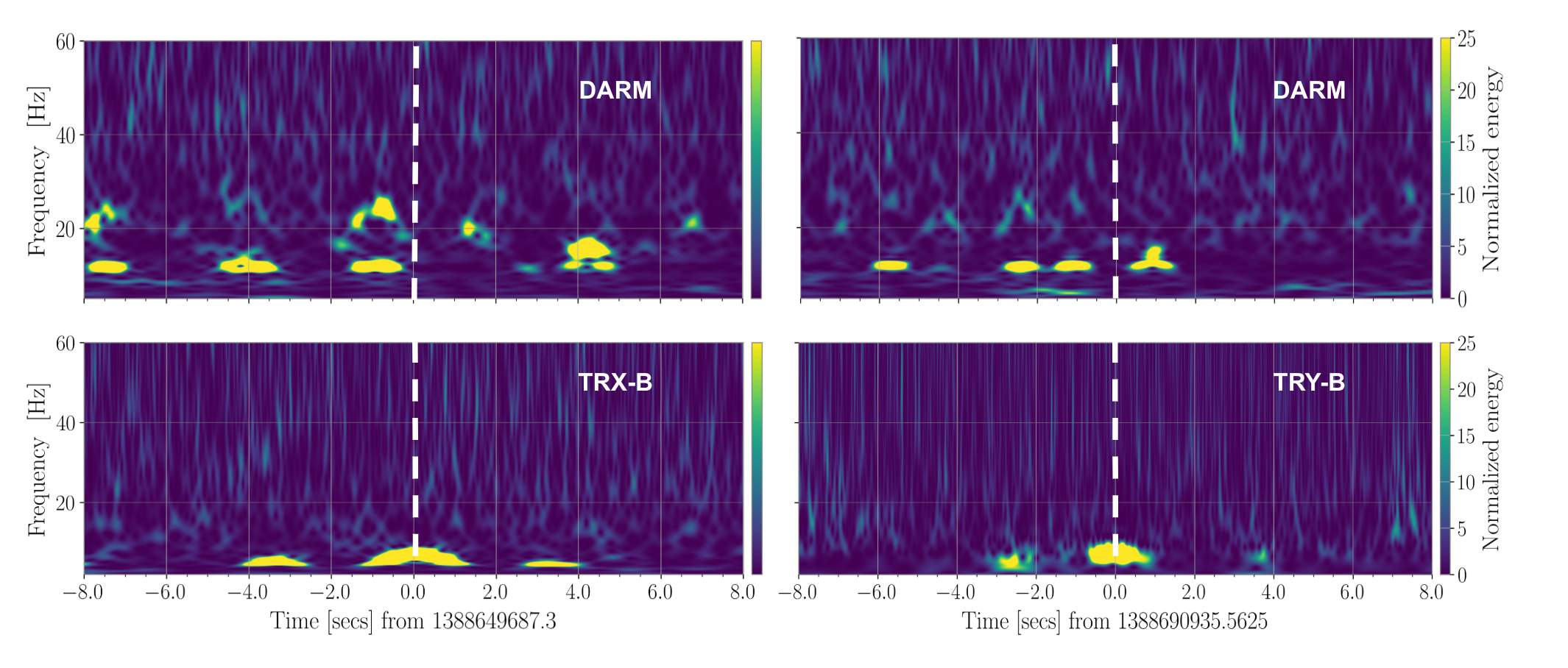}
    \caption{\textit{Top}: Glitches in DARM produced by scattered light; \textit{Bottom}: Glitches in the photodiodes of X-end (TRX) and Y-end (TRY) transmitted light monitors. The dashed white line is at $t=0$, at this time there are glitches in the photodiodes of the transmitted light monitors but not in DARM.}
    \label{fig:transmon_scatter}
\end{figure}

A small amount of light is being scattered between the ETMs and the transmitted light monitors, which are just behind the ETM mirrors. During O3, the control drive was applied only at ETMX and it induced a high amount of motion at the test mass mirror. This large relative motion between the ETMX mirror and the transmitted light monitor was responsible for producing some scattered light noise. In O4, this control drive was split between the two end stations, so the motion between the ETMX mirror and the transmitted light monitor is half of what it was before. Still, since the detector has a better sensitivity now, it is possible that the noise produced by that much motion might show up above the present noise background of the detector.

\begin{figure*}
    \centering
    \includegraphics[width=0.85\textwidth]{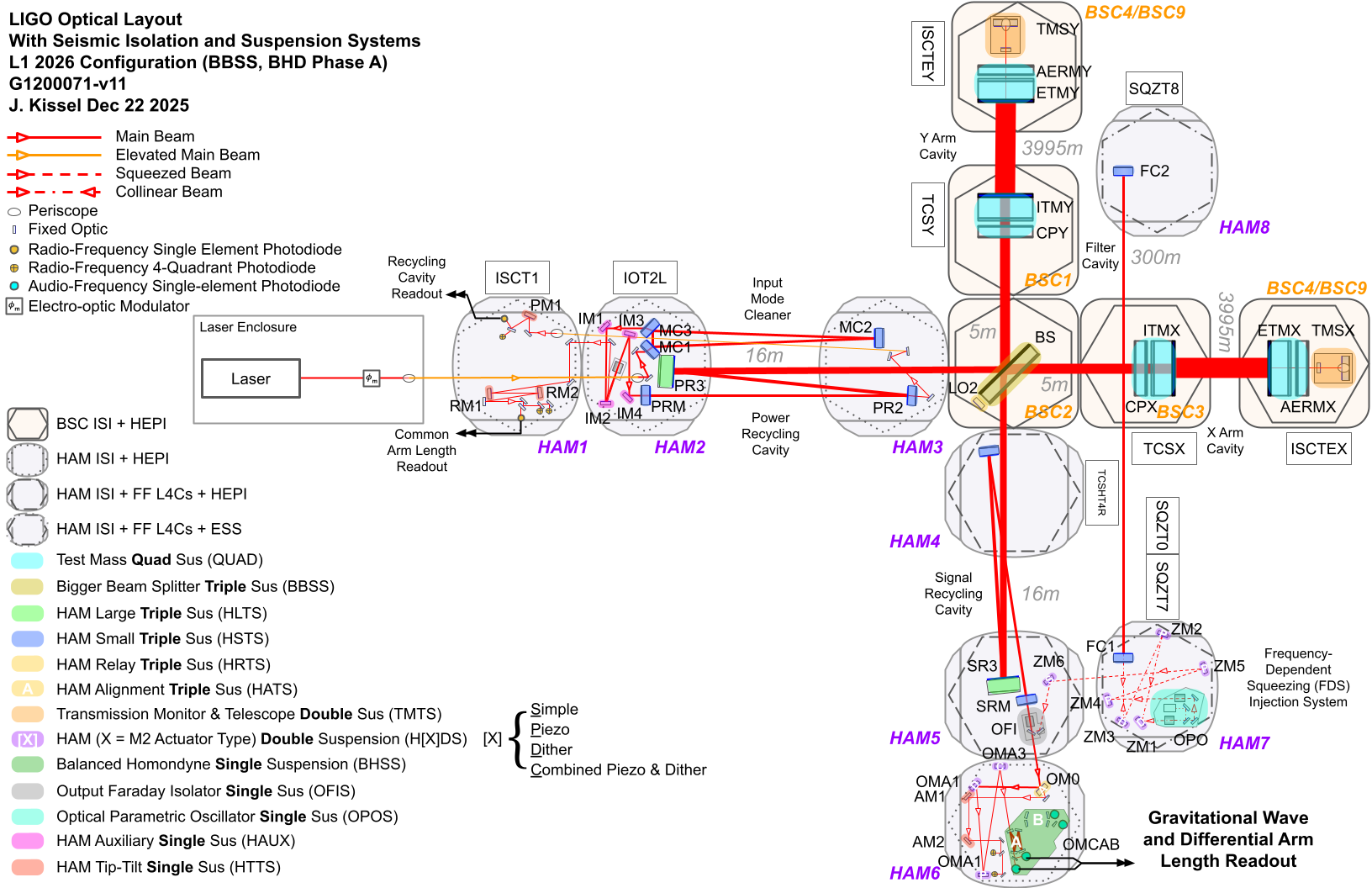}
    \caption{Optical layout of LLO}
    \label{fig:layout}
\end{figure*}

We checked the four photodiodes (2 at each end station, namely TRX-A/B and TRY-A/B) at the transmitted light monitors of both the end stations on a few days of O4 when the microseismic ground motion was very high. Even with a very high amount of ground motion, three of these photodiodes (TRX-A/B and TRY-B) had a negligible amount of glitches between $10$-$60$ Hz, which is typically where scattered light glitches appear. One of the photodiodes at the Y-end transmitted light monitor (TRY-A) had a significant number of glitches present between $10$-$60$ Hz, but after doing the hierarchical veto (hveto) analysis \cite{Smith:2011an}, these glitches were proved not to be coincident with any of the glitches present in DARM. Figure \ref{fig:transmon_scatter} shows an example where we see glitches in the X- and Y-end transmitted light monitors, but these glitches are not coincident with the ones present in DARM at that time and also their frequencies are lower than those which appear in DARM.

% The \nocite command causes all entries in a bibliography to be printed out
% whether or not they are actually referenced in the text. This is appropriate
% for the sample file to show the different styles of references, but authors
% most likely will not want to use it.
%\nocite{*}

\clearpage
\bibliography{apssamp}% Produces the bibliography via BibTeX.

\end{document}